\documentclass{article}
\PassOptionsToPackage{numbers,sort&compress}{natbib}
\usepackage[main,final]{neurips_2026}

\usepackage[utf8]{inputenc}
\usepackage[T1]{fontenc}
\usepackage{courier}
\usepackage{hyperref}
\usepackage{url}
\usepackage{booktabs}
\usepackage{amsfonts}
\usepackage{amsmath}
\usepackage{amssymb}
\usepackage{nicefrac}
\usepackage{microtype}
\usepackage[table]{xcolor}
\usepackage{graphicx}
\usepackage{subcaption}
\usepackage{enumitem}
\usepackage{multirow}
\usepackage{tabularx}
\usepackage{placeins}

\usepackage{amsthm}

\definecolor{canyon}{RGB}{205, 92, 92}

\title{AlloGen: Conformation-Selective Binder Design with Differential State Scoring}

\author{%
  Hanqun Cao$^{1}$ \quad Aastha Pal$^{2}$ \quad Sumi Kimura$^{2}$ \quad Yesol Kim$^{2}$ \\
  \textbf{Jingjie Zhang}$^{1}$ \quad \textbf{Pheng Ann Heng}$^{1}$ \quad \textbf{Pranam Chatterjee}$^{2,3,\dagger}$ \\
  \\[0.4em]
  $^{1}$Department of Computer Science and Engineering,
  The Chinese University of Hong Kong \\
  $^{2}$Department of Bioengineering,
  University of Pennsylvania \\
  $^{3}$Department of Computer and Information Science,
  University of Pennsylvania \\[0.4em]
  $^{\dagger}$Corresponding Author:
  \href{mailto:pranam@upenn.edu}
       {\texttt{pranam@upenn.edu}} \\
  Code: \url{https://huggingface.co/ChatterjeeLab/AlloGen}
}

\hypersetup{hidelinks,
  pdftitle={AlloGen: Conformation-Selective Binder Design with Differential State Scoring},
  pdfauthor={Hanqun Cao, Aastha Pal, Sumi Kimura, Yesol Kim, Jingjie Zhang, Pheng-Ann Heng, Pranam Chatterjee}}
\begin{document}

\maketitle
\begin{abstract}
Protein binder design commonly optimizes compatibility with a single receptor conformation, leaving preference over alternative functional states unresolved. To address this gap, we introduce \textbf{AlloGen}, a modular method that decouples backbone generation from state-selectivity scoring. We train its SE(3)-invariant interface graph transformer, $Q_\theta$, through interface-quality regression followed by paired conformational discrimination. By comparing scores across states, we rerank candidates and guide frozen backbone generators without retraining. On a benchmark of 65 targets spanning 15 protein families, $Q_\theta$ achieved a mean Spearman correlation of $0.520$ on eight held-out targets. All 15 evaluated generator--guidance combinations achieved positive predicted selectivity averaged over these targets, and RFdiffusion with Langevin refinement reached $\bar{S}_{\mathrm{cons}}=0.677$ on calmodulin (CaM). To test the predicted preference experimentally, we measured binding by bio-layer interferometry. Five designed peptides bound holo CaM, with the strongest measured $K_D$ of $46.6$\,nM, and none showed detectable apo CaM binding under the tested conditions. Together, these results support computational prioritization of conformation-selective binders.
\end{abstract}

\begin{figure}[htbp]
\centering
\includegraphics[width=\linewidth]{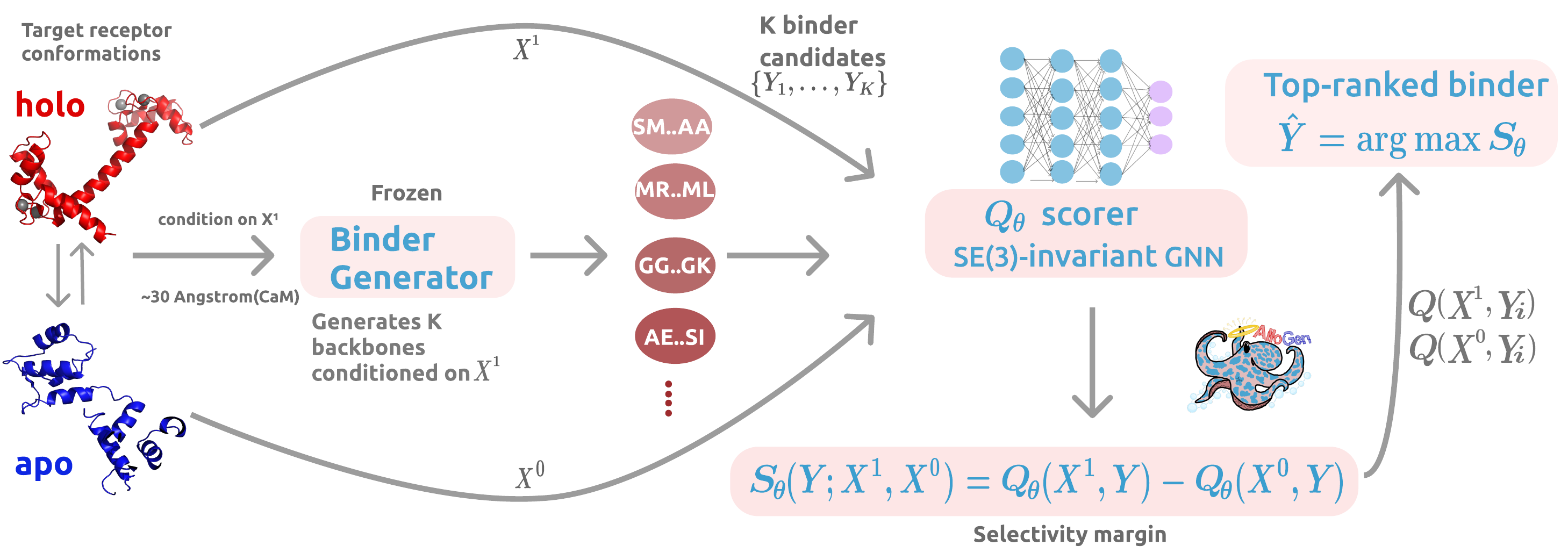}
\caption{\textbf{AlloGen pipeline.}
(\textit{Left}) We use two receptor conformations: the goal state $X^1$ (holo, red) and the undesired state $X^0$ (apo, blue). (\textit{Centre}) A frozen pretrained generator produces $K$ binder backbone candidates conditioned on $X^1$. (\textit{Right}) We score each candidate in both conformational frames with $Q_\theta$ and select the top candidate $\hat{Y}$ by the selectivity gap $\Delta Q = Q_\theta(X^1,Y) - Q_\theta(X^0,Y)$ through best-of-$K$ reranking. DockQ regression followed by contrastive fine-tuning trains $Q_\theta$ independently of the backbone generator, enabling reuse across generators without retraining.}
\label{fig:AlloGen_frame1}
\end{figure}

\section{Introduction}
\label{sec:introduction}
Protein conformational transitions govern signaling, catalysis, and regulation across diverse protein families~\citep{vetter2003novel, ha2012protein, weikl2014conformational, nussinov2016introduction}. In kinases, nuclear receptors, and GPCRs, these transitions connect conformations with distinct biological functions. Binder design therefore requires control over conformational preference as well as binding affinity~\citep{kar2010allostery, vijayan2015conformational, kojetin2013small, conflitti2025functional}. This capability is relevant to allosteric drug design, conformational biosensors, and synthetic biology switches~\citep{kar2010allostery, langan2019novo}.

Generative models support protein binder design at both sequence and structural levels. At the sequence level, masked language models generate peptide binders conditioned on target sequence~\citep{chen2025target}, contrastive language models enable de novo peptide design for conformationally diverse targets~\citep{bhat2025novo}, and multi-objective discrete diffusion supports therapeutic peptide generation~\citep{tang2025peptune, vincoff2025soapia, cao2025glide, cao2026tdb}. Structure-conditioned methods extend these capabilities to binder geometry. RFdiffusion generates functional protein binders~\citep{watson2023novo}, PXDesign supports fast, modular binder design~\citep{team2025pxdesign}, and Proteina-ComplexA enables atomistic design through generative pretraining and test-time compute~\citep{didi2026scaling}. BoltzGen supports binder design across protein families~\citep{stark2025boltzgen}, and BindCraft combines structure prediction with sequence optimization to produce experimentally validated binders~\citep{pacesa2025one}. In standard single-state settings, these workflows optimize compatibility with one receptor conformation, although a resulting binder may also bind an alternative state. State-selective design therefore requires an explicit comparison across conformations, which single-state interface scores can assess only through separate evaluations.

To address this gap, we introduce \textbf{AlloGen}, which decouples binder generation from selectivity evaluation (Figure~\ref{fig:AlloGen_frame1}). We train its SE(3)-invariant interface graph transformer, $Q_\theta$, on paired receptor conformations to score each candidate in both states. The difference between these scores then guides candidate reranking, gradient-based refinement, and resampling with frozen generators~\citep{dhariwal2021diffusion, gruver2023protein}. This separation allows us to apply the same scorer across generator architectures without retraining.

\textbf{Our contributions are as follows:}
\begin{enumerate}[leftmargin=*,topsep=2pt,itemsep=1pt]
    \item \textbf{Problem formulation, benchmark, and scorer.} We develop a paired-state scoring task with: 
    \textbf{(1)} a benchmark of 65 two-state targets across 15 protein families and 2{,}896 receptor--binder complexes covering both apo and holo conformations;
    \textbf{(2)} $Q_\theta$, an SE(3)-invariant interface graph transformer that scores backbone-only candidates against paired receptor conformations; and 
    \textbf{(3)} a two-phase training curriculum for interface-quality regression followed by conformational discrimination. On eight held-out targets, $Q_\theta$ achieved $\bar{\rho}=0.520$, exceeding the evaluated contact-based proxies.
    \item \textbf{Scoring controls and experimental validation.} We assess target specificity and apo rejection in the main text (Section~\ref{sec:exp_conformational}), then compare selectivity scores with four $Q_\theta$-independent metrics in the appendix: ProteinMPNN log-likelihood, Boltz-2 interface predicted TM-score (ipTM), AlphaFold~3 ipTM, and Rosetta InterfaceAnalyzer. BLI measurements on CaM directly test state-dependent binding in selected designs.
    \item \textbf{End-to-end selectivity-guided design.} All 15 generator--guidance combinations achieved positive mean predicted selectivity over the eight test targets. TDS and SMC had the highest cross-target means for each generator. In a CaM case study, RFdiffusion combined with Langevin refinement reached $\bar{S}_{\mathrm{cons}}=+0.677$ at $88\%$ design success rate, and best-of-$K$ reranking reached $\bar{S}=+0.885$ at $K=10$. Five selected peptides showed measurable holo CaM binding with no detectable apo binding (Section~\ref{sec:exp_wetlab}).
\end{enumerate}

\section{Related Work}
\label{sec:relatedwork}
\paragraph{Structure-Based Binder Design.}
\label{sec:related_binder}
Structure-based generative models design binders conditioned on target geometry. RFdiffusion~\citep{watson2023novo} generates binder backbones conditioned on a fixed target surface via backbone diffusion, followed by sequence design with ProteinMPNN~\citep{dauparas2022robust}. PPDiff~\citep{song2025ppdiff} and PXDesign~\citep{team2025pxdesign} extend this to joint sequence-structure generation, while Proteina-ComplexA~\citep{didi2026scaling} employs atomistic flow matching to produce full-atom binder complexes. On the sequence side, a growing body of work generates binder sequences through discrete diffusion and language model objectives~\citep{chen2025target, bhat2025novo, chen2025moppit, vincoff2025soapia, tang2025peptune, tang2025tr2, chen2025areuredi, tang2025gumbel, cao2025glide, cao2026tdb}. Standard structure-conditioned design settings optimize compatibility with one receptor conformation. AlloGen adds an explicit two-state score for reranking or guiding their backbone outputs.

\paragraph{Protein--Protein Interface Scoring.}
\label{sec:related_scoring}
Physics-based methods such as FoldX~\citep{schymkowitz2005foldx} and PRODIGY~\citep{xue2016prodigy} estimate binding free energies, and inverse folding models such as ProteinMPNN~\citep{dauparas2022robust} and ESM-IF~\citep{hsu2022learning} serve as zero-shot affinity proxies. GNN-based scorers including GNN-DOVE~\citep{wang2021protein} and DeepRank-GNN-esm~\citep{xu2024deeprank} learn interface quality from structural features for docking re-ranking. Comparing such scores across receptor states assesses conformational preference. To train explicitly for this comparison, we combine interface-quality regression with paired contrastive fine-tuning of $Q_\theta$.

\paragraph{Conformational Selectivity: Sampling, Design, and Negative Design.}
\label{sec:related_selectivity}
Conformation-selective design requires structural characterization of alternative states and binders that discriminate between them. On the characterization side, AlphaFlow~\citep{jing2024alphafold} uses flow matching to generate structural ensembles, AFsample2~\citep{kalakoti2025afsample2} diversifies predictions via MSA subsampling, and AlphaFold~3~\citep{abramson2024accurate} extends structure prediction to biomolecular complexes. Designing binders that distinguish these states further requires objectives for molecular recognition. In this setting, negative design has a long history in computational protein engineering~\citep{stranges2013comparison, fleishman2011computational}, and multi-state design in Rosetta~\citep{havranek2003automated} optimizes sequences across multiple conformations. Extending these sequence-level objectives to backbone generation requires a differentiable score for receptor--binder geometry. Contrastive objectives inspired by InfoNCE~\citep{oord2018representation} have been applied to molecular representation learning, and classifier guidance~\citep{dhariwal2021diffusion} and TDS~\citep{wu2023practical} support gradient-based steering and particle reweighting, respectively. Concurrent work by \citet{xue2025state} targets state-specific peptide design for GPCRs via a conformation-specific folding filter, and ProDiT~\citep{jing2025generating} introduces multi-state structure diffusion. AlloGen combines paired contrastive scoring with guidance of three generators and evaluates transfer on eight held-out targets from a 65-target benchmark spanning 15 families.

\section{Preliminaries}
\label{sec:preliminaries}
\subsection{Problem Formulation}
\label{sec:prelim_problem}
We consider a target protein that exists in two distinct conformational states: an \emph{apo} (undesired) conformation $X^0$ and a \emph{holo} (goal) conformation $X^1$, both represented as backbone coordinate sets. Given a binder $Y$ represented by its backbone coordinates, the \emph{state-selectivity scoring problem} is to learn a function $Q_\theta : \mathcal{X} \times \mathcal{Y} \to (0, 1)$ such that for an $X^1$-selective candidate $Y$:
\begin{align}
  Q_\theta(X^1, Y) \gg Q_\theta(X^0, Y), \label{eq:selectivity_objective}
\end{align}
while non-selective binders, apo-preferring binders, and non-binders are not required to satisfy this inequality. The training objective in Section~\ref{sec:method_training} enforces this conditional behaviour through paired contrastive supervision.
Given a pretrained binder generator $p_\psi(Y \mid X^1)$, the \emph{two-state binder design} task is to identify the most conformationally selective candidate $\hat{Y} = \arg\max_{Y^{(k)}} S_\theta(Y^{(k)}; X^1, \mathcal{N})$ over $K$ samples $Y^{(k)} \sim p_\psi(\cdot \mid X^1)$, where $S_\theta$ is the selectivity margin defined in Section~\ref{sec:method_inference} and $\mathcal{N} = \{X^0\}$ is the set of undesired conformations.

\subsection{Protein Backbone Representation and SE(3) Invariance}
\label{sec:prelim_backbone}
We represent each residue $i$ by its $\mathrm{C}_\alpha$ position $\mathbf{p}_i \in \mathbb{R}^3$ and a local backbone frame $R_i \in \mathrm{SO}(3)$ constructed from the $(\mathrm{N}, \mathrm{C}_\alpha, \mathrm{C})$ triplet via Gram--Schmidt orthogonalization~\citep{jumper2021highly}. While $\mathbf{p}_i$ and $R_i$ themselves transform with global rigid motions, the pair $(\mathbf{p}_i, R_i)$ defines a residue-local frame from which all inter-residue geometry can be expressed in an SE(3)-invariant way. For $Q_\theta$ to be physically meaningful, it must be SE(3)-invariant: $Q_\theta(gX, gY) = Q_\theta(X, Y)$ for all rigid motions $g \in \mathrm{SE}(3)$, ensuring scores are independent of the global position and orientation of the complex. We enforce this by expressing all inter-residue geometry in the local frame of residue $i$: distances $\|\mathbf{p}_j - \mathbf{p}_i\|$, directions $R_i^\top(\mathbf{p}_j - \mathbf{p}_i)/\|\cdot\|$, and relative orientations $R_i^\top R_j$. Each of these quantities is invariant under rigid motions $g$ applied jointly to both $X$ and $Y$, so the resulting node and edge features, and hence $Q_\theta$, are SE(3)-invariant.

\subsection{DockQ as Interface Quality Proxy}
\label{sec:prelim_dockq}
DockQ~\citep{basu2016dockq} is a composite scalar $\in [0,1]$ that measures protein--protein docking quality by combining the fraction of native contacts, interface RMSD, and ligand RMSD. We use this geometric measure to supervise Phase~1 training (Section~\ref{sec:method_training}), initializing the scorer to assess receptor--binder fit before paired conformational discrimination in Phase~2. A DockQ value $> 0.23$ is the conventional threshold for an acceptable docking model~\citep{basu2016dockq}.

\section{Method}
\label{sec:method}
\subsection{Interface Graph Construction}
\label{sec:method_graph}

To score local receptor--binder geometry, we construct a sparse interface graph $\mathcal{G} = (\mathcal{V}, \mathcal{E})$ for each complex $(X, Y)$. The node set $\mathcal{V}$ contains residues from both $X$ and $Y$ with at least one inter-chain $\mathrm{C}_\alpha$ contact within an 8\,\AA{} cutoff, and edges connect all residue pairs within that cutoff.

\paragraph{Node features.}
Each node $\mathbf{h}_i$ encodes four types of information. \emph{Amino acid identity} is represented as a one-hot vector over 20 standard amino acids plus an unknown token, providing residue-level sequence identity. \emph{Backbone torsion angles} $\varphi$, $\psi$, and $\omega$ are encoded via their sine and cosine values, yielding a smooth periodic representation of local backbone conformation. \emph{Sidechain torsion angles} $\chi_1$ and $\chi_2$ are similarly encoded to capture rotameric states at the interface; residues without sidechain degrees of freedom are zero-padded. A \emph{chain indicator} flag distinguishes receptor from binder residues. When available, per-residue ESM-2 embeddings~\citep{lin2023evolutionary} are projected and concatenated to these structural features, providing evolutionary context beyond local geometry.

\paragraph{Edge features.}
All edge features $\mathbf{e}_{ij}$ are computed in the local backbone frame of residue $i$, ensuring SE(3) invariance. \emph{Geometric features} encode the inter-residue distance via a Gaussian RBF basis, the unit direction $R_i^\top(\mathbf{p}_j - \mathbf{p}_i)/\|\mathbf{p}_j - \mathbf{p}_i\|$, and the relative backbone orientation $R_i^\top R_j$, jointly capturing the full relative rigid-body relationship between residues. \emph{Sequence separation} is encoded as a binned index difference within a chain and set to the maximum bin for inter-chain pairs, allowing the model to distinguish intra- from inter-chain interactions. A \emph{same-chain indicator} flag further disambiguates receptor-receptor, binder-binder, and receptor-binder edges.

\subsection{State-Selectivity Scorer \texorpdfstring{$Q_\theta$}{Q\_theta}}
\label{sec:method_scorer}

Using these invariant features, we assess receptor--binder fit with a dense edge-biased graph transformer~\citep{vaswani2017attention}, $Q_\theta$. Comparing its outputs across conformations yields the selectivity score. Given node embeddings $\mathbf{H}^{(0)}$ and edge embeddings $\mathbf{E}^{(0)}$, the model applies $L = 4$ transformer layers. At each layer $\ell$, attention weights are computed as:
\begin{align}
  \alpha_{ij}^{(\ell)} &= \frac{(\mathbf{h}_i^{(\ell)} W_Q)(\mathbf{h}_j^{(\ell)} W_K)^\top}{\sqrt{d_h}} + b_{ij}, \quad b_{ij} = \mathbf{e}_{ij} W_E \in \mathbb{R}, \label{eq:attention}\\
  \mathbf{h}_i^{(\ell+1)} &= \mathbf{h}_i^{(\ell)} + \mathrm{FFN}\!\left(\sum_j \mathrm{softmax}_j(\alpha_{ij}^{(\ell)}) \cdot \mathbf{h}_j^{(\ell)} W_V\right), \label{eq:update}
\end{align}
where $W_E \in \mathbb{R}^{d_e \times 1}$ projects each edge embedding to a per-head scalar attention bias $b_{ij}$ that is added directly to the dot-product logit (one such projection is learned per attention head). Edge embeddings are computed once and shared across all layers. After $L$ layers, mean- and max-pooled node representations are concatenated to form $\mathbf{h}_{\mathrm{pool}}^{(L)} = [\mathrm{mean}_i\, \mathbf{h}_i^{(L)};\, \mathrm{max}_i\, \mathbf{h}_i^{(L)}]$ and passed through an MLP with sigmoid activation to produce:
\begin{equation}
  Q_\theta(X, Y) = \sigma(\mathrm{MLP}(\mathbf{h}_{\mathrm{pool}}^{(L)})) \in (0, 1). \label{eq:qtheta}
\end{equation}
The bounded output yields a natural selectivity gap $Q_\theta(X^1, Y) - Q_\theta(X^0, Y) \in (-1, 1)$. Architectural hyperparameters and parameter count are reported in Appendix~\ref{app:implementation}.

\subsection{Two-Phase Training of \texorpdfstring{$Q_\theta$}{Q\_theta}}
\label{sec:method_training}

To learn both interface quality and conformational preference, we train $Q_\theta$ in two phases. We first establish a representation of receptor--binder geometry, then optimize paired state discrimination.

\paragraph{Phase 1: Interface Quality Regression.}
We first train $Q_\theta$ to predict interface quality as measured by DockQ~\citep{basu2016dockq}:
\begin{equation}
  \mathcal{L}_q = \mathrm{MSE}(Q_\theta(X, Y),\; d_{\mathrm{DockQ}}(X, Y)). \label{eq:phase1_loss}
\end{equation}
The training data include native holo complexes, apo mismatches, rigid-body decoys, and hard negatives described in Section~\ref{sec:exp_setup}. Together, these examples initialize the scorer to distinguish compatible interfaces from mismatches before Phase~2.

\paragraph{Phase 2: Selectivity Fine-Tuning.}
Building on this initialization, we fine-tune $Q_\theta$ to distinguish conformational preference from target-specific biases. For paired triplets $(X^+, X^-, Y)$ where $X^+ = X^1$ and $X^- = X^0$ for the same binder $Y$, we minimize a multi-negative InfoNCE loss~\citep{oord2018representation} that includes both apo negatives (the same binder against its target's apo conformation) and cross-target negatives (the anchor's holo conformation paired with binders from other targets in the batch). The apo term trains conformational discrimination, while the cross-target term discourages high scores based on receptor identity alone.
\begin{align}
  \mathcal{L}_{\mathrm{NCE}} &= -\frac{1}{B}\sum_{i=1}^{B} \log \frac{\exp(Q_\theta(X^+_i, Y_i) / \tau)}{Z_i}, \label{eq:nce_loss} \\
  Z_i = \exp(Q_\theta(X^+_i, Y_i)/\tau) & + \exp(Q_\theta(X^-_i, Y_i)/\tau) \nonumber + \sum_{\substack{k=1 \\ k \neq i}}^{B} \exp(Q_\theta(X^+_i, Y_k)/\tau), \nonumber
\end{align}
where $\tau$ is a temperature hyperparameter reported in Appendix~\ref{app:implementation}. The regression loss $\mathcal{L}_q$ is dropped in Phase~2 to avoid conflicting gradient signals; the curriculum comparison is reported in Figure~\ref{fig:ablation_train_feat}.

\paragraph{Backbone-geometry augmentation.}
Training complexes have known sequences, but generated binder backbones lack sequence identity and sidechain coordinates. To train $Q_\theta$ for this backbone-only inference setting, we mask binder-side sequence features with probability $p_{\mathrm{drop}}$:
\begin{equation}
  \tilde{\mathbf{h}}_i^{\mathrm{bnd}} = \bigl[\mathbf{0}_{\mathrm{AA}},\; \sin\varphi_i,\, \cos\varphi_i,\, \sin\psi_i,\, \cos\psi_i, \sin\omega_i,\, \cos\omega_i,\; \mathbf{0}_{\chi},\; \mathbf{0}_{\mathrm{ESM}},\; 1\bigr], \label{eq:binder_mask}
\end{equation}
where $\mathbf{0}_{\mathrm{AA}}$, $\mathbf{0}_{\chi}$, and $\mathbf{0}_{\mathrm{ESM}}$ replace the amino acid identity, sidechain torsion angles, and ESM-2 embeddings with zeros, respectively. All backbone torsions and edge features are preserved, so scoring can use backbone geometry when binder sequence features are unavailable. The same masking decision is applied consistently to both $X^+$ and $X^-$ within each training pair.

\subsection{Selectivity-Guided Binder Design}
\label{sec:method_inference}

To use the trained scorer for design, we sample $K$ binder candidates $Y^{(1)}, \ldots, Y^{(K)}$ from the frozen generator $p_\psi(\cdot \mid X^1)$ and evaluate each against $X^1$ and $X^0$ by rigid placement onto their aligned structures. We define the \emph{selectivity margin} as:
\begin{equation}
  S_\theta(Y;\, X^1, \mathcal{N}) = \mathrm{logit}(Q_\theta(X^1, Y)) - \log \sum_{X^- \in \mathcal{N}} \exp(\mathrm{logit}(Q_\theta(X^-, Y))), \label{eq:selectivity_margin}
\end{equation}
where $\mathcal{N} = \{X^0\}$ is the set of undesired conformations. $S_\theta$ is positive when the holo score exceeds the apo score, zero when the two scores are equal, and negative when the apo score is higher. The logit-space formulation extends naturally to multi-state targets with $|\mathcal{N}| > 1$. Candidates are filtered by minimum interface size and steric clash criteria before selection, and sequence design is performed with ProteinMPNN~\citep{dauparas2022robust} on the selected backbone.

\paragraph{Gradient-based guidance.}
Beyond reranking, we use four guidance strategies to update coordinates or reweight candidates with the trained scorer. \emph{Langevin backbone refinement}~\citep{song2021scorebased,welling2011bayesian} performs deterministic gradient ascent on $S_\theta$ for a completed backbone: $\mathbf{x}_{t+1} = \mathbf{x}_t + \eta\,\nabla_{\mathbf{x}} S_\theta(\mathbf{x}_t;\, X^1, \mathcal{N})$, where $\eta$ is the step size; this deterministic variant uses no injected Langevin noise and operates on fully denoised backbones. \emph{Classifier guidance}~\citep{dhariwal2021diffusion} uses holo-score gradients during denoising to favor holo compatibility; gradient reliability decreases at high noise levels. \emph{Twisted diffusion sampling} (TDS)~\citep{wu2023practical} reweights diffusion particles using holo scores without gradient-based coordinate updates. \emph{Sequential Monte Carlo} (SMC)~\citep{wu2023practical} iteratively resamples complete trajectories by $S_\theta$ across multiple generation rounds, enriching the candidate pool for predicted selectivity. Detailed formulations and computational cost are provided in Appendix~\ref{app:guidance_formulations} and~\ref{app:efficiency}.

\section{Experiments}
\label{sec:experiments}

We first tested interface-quality ranking and conformational discrimination on held-out targets (Sections~\ref{sec:exp_scoring}--\ref{sec:exp_conformational}), then evaluated the scorer for binder design (Section~\ref{sec:exp_design}) and measured state-dependent binding in selected CaM designs (Section~\ref{sec:exp_wetlab}).

\subsection{Experimental Setup}
\label{sec:exp_setup}

\paragraph{Target selection.}
We curated 65 two-state proteins spanning 15 protein families and diverse conformational mechanisms (Table~\ref{tab:expanded_families}). We required: (1) experimentally determined structures for both apo and holo conformational states, (2) at least three co-crystal structures with peptide or protein binders in the goal state, and (3) a structurally defined conformational change between states. Targets range from large-scale domain rearrangements (\textbf{CaM}: ${\sim}30$\,\AA\ apo-to-holo; \textbf{ABL1}: DFG-loop flip, ${\sim}6.5$\,\AA\ global) to subtle helix repositioning (\textbf{ER$\alpha$}: H12 ${\sim}10$\,\AA; \textbf{CDK2}: Cyclin~A-induced). The dataset spans kinases (9), small GTPases (6), nuclear receptors (5), GPCRs and ion channels (6), proteases (6), and the remaining families, totalling 2{,}896 complexes across 65 targets.

\paragraph{Sample construction.}
Each complex yields 12 base training samples: 1 positive native holo complex $(X^1, Y^{\mathrm{native}})$ with label $1.0$; 1 negative apo-mismatch complex $(X^0_{\mathrm{aligned}}, Y^{\mathrm{native}})$ with label $0.0$, where the apo receptor is Kabsch-aligned to the holo frame; and 10 rigid-body decoys at target C$_\alpha$ RMSD levels spanning 1--8\,\AA, with labels $\max(0, 1 - d_{\mathrm{RMSD}}/4)$. Prior to augmentation, \textbf{FastRelax Neg.}\ are generated by constrained relaxation of binders placed on apo receptors, producing 959 hard negatives included in all configurations. Three augmentation types further enrich training: \textbf{Cross-family negatives} (Cross-fam) place native binders on structurally unrelated receptors; \textbf{conformational decoys} (Conf.\ decoys) apply Rosetta FastRelax repacking on the holo receptor to produce near-native hard negatives; and \textbf{generator decoys} (GenDecoys) are synthetic binders from structure-based generative models that provide diverse non-native interface geometries.

\paragraph{Dataset splits and configurations.}
All main results use the \textbf{target split}, in which targets are partitioned 51/6/8 across train, OOD validation, and OOD test, with CaM held out as the primary OOD design target (Table~\ref{tab:split_v5}). The \textbf{Baseline} dataset comprises 51 training, 6 OOD validation (14-3-3, 14-3-3$\sigma$, $\beta_2$AR, Caspase-3, $\mu$-opioid, Rac1), and 8 OOD test targets (CaM, BCL-2, ER$\alpha$, MDM2, Ran, A$_{2A}$, PAI-1, Integrin); the \textbf{Augmented} dataset extends it with GenDecoys and cross-family negatives generated exclusively on training-set targets, excluding test-target generator decoys from training. To isolate augmentation sources, we define three scoring configurations (Table~\ref{tab:dataset_versions}): \textbf{S1} uses Baseline only, \textbf{S2} uses full Augmented data, and \textbf{S3} adds Cross-fam and Conf.\ decoys, withholding GenDecoys.

\paragraph{Metrics.}
We report Spearman $\rho$ with DockQ and the bounded score gap $S(Y)=Q_\theta(X^1,Y)-Q_\theta(X^0,Y)$. Tables and plots use this bounded gap; $S_\theta$ in Eq.~\eqref{eq:selectivity_margin} denotes the distinct logit-space objective. Consensus selectivity $\bar{S}_{\mathrm{cons}}$ averages the bounded gap across designs and three independently trained checkpoints. We use ProteinMPNN $\Delta$NLL, Boltz-2 $\Delta$ipTM, and AlphaFold~3 $\Delta$ipTM as independent computational measures of state preference. Implementation details are in Appendix~\ref{app:implementation}.

\begin{table}[htbp]
\centering
\caption{Energy-based baselines vs.\ $Q_\theta$ on 8 OOD targets. Spearman $\rho$.}
\label{tab:energy_baselines}
\begin{tabular}{l r c c c c c}
\toprule
\textbf{Target} & $n$ & PRODIGY & Iface Size & Edge Dens. & Random & $Q_\theta$ \\
\midrule
A$_{2A}$R  &   36 & $-0.230$ & $-0.119$ & $-0.092$ & $-0.120$ & $\mathbf{0.469} \pm .034$ \\
BCL-2      &  132 & $+0.160$ & $-0.050$ & $-0.082$ & $-0.039$ & $\mathbf{0.667} \pm .014$ \\
CaM        &   96 & $+0.199$ & $-0.316$ & $+0.014$ & $+0.059$ & $\mathbf{0.474} \pm .061$ \\
ER$\alpha$ &   72 & $+0.551$ & $-0.292$ & $-0.036$ & $+0.086$ & $\mathbf{0.664} \pm .013$ \\
Integrin   &   60 & $-0.019$ & $-0.195$ & $+0.053$ & $+0.076$ & $\mathbf{0.366} \pm .085$ \\
MDM2       &  143 & $+0.163$ & $-0.050$ & $-0.130$ & $-0.037$ & $\mathbf{0.589} \pm .036$ \\
PAI-1      &  156 & $+0.058$ & $+0.062$ & $-0.146$ & $-0.043$ & $\mathbf{0.421} \pm .045$ \\
Ran        & 2268 & $+0.264$ & $-0.270$ & $-0.138$ & $-0.018$ & $\mathbf{0.511} \pm .065$ \\
\midrule
\textbf{Mean} & & $+0.143$ & $-0.154$ & $-0.070$ & $-0.005$ & $\mathbf{0.520} \pm .010$ \\
\bottomrule
\end{tabular}
\end{table}

\subsection{Scorer Performance}
\label{sec:exp_scoring}
\begin{figure*}[htbp]
\centering
\includegraphics[width=\linewidth]{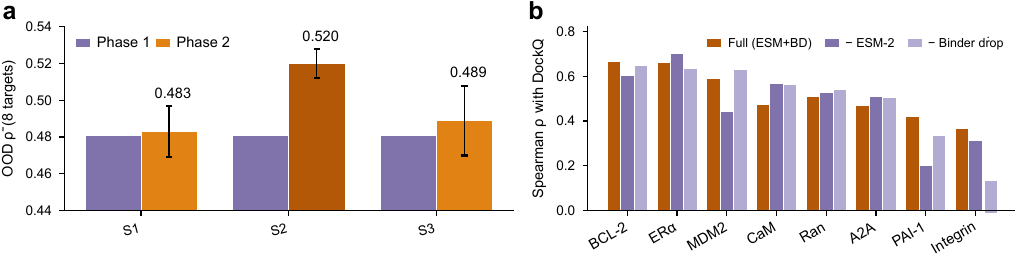}
\caption{\textbf{$Q_\theta$ selectivity performance.}
\textbf{(a)} $Q_\theta$ scoring performance ablation by data augmentation strategies and two phases; \textbf{(b)} $Q_\theta$ scoring performance ablation per target by different features.}
\label{fig:ablation_train_feat}
\end{figure*}

To test transfer to unseen targets, we evaluated $Q_\theta$ on the eight held-out test targets. $Q_\theta$ attained $\bar\rho{=}\mathbf{0.520 \pm 0.010}$ (3 random seeds for training), with positive correlation on all 8 OOD targets and 4 of 8 exceeding $\rho{=}0.5$ (Figure~\ref{fig:ablation_train_feat}). 

\paragraph{Training curriculum, data augmentation, and feature design.}
We then used ablations to assess the contributions of training, augmentation, and features (Figure~\ref{fig:ablation_train_feat}). \emph{(i) Two-phase curriculum.} Phase~2 contrastive fine-tuning improved over Phase~1 regression across all augmentation configurations (Figure~\ref{fig:ablation_train_feat}a). \emph{(ii) Data augmentation.} The full Augmented configuration yielded the largest single gain ($\Delta\bar\rho{=}{+}0.037$), driven by GenDecoys, whose synthetic binder geometries broadened interface coverage beyond rigid-body and FastRelax decoys. \emph{(iii) Feature design.} In per-target ablations (Figure~\ref{fig:ablation_train_feat}b), removing ESM-2 reduced $\bar\rho$ across most targets, and disabling binder dropout further reduced performance, with the largest drops on harder OOD targets (Integrin, PAI-1, A$_{2A}$). The combined Full configuration (ESM\,+\,BD) performed best on all 8 OOD targets. InfoNCE batch-size ablations are in Appendix~\ref{app:scorer_details}, Table~\ref{tab:phase2_ablation}.

\begin{figure*}[htbp]
\centering
\includegraphics[width=\linewidth]{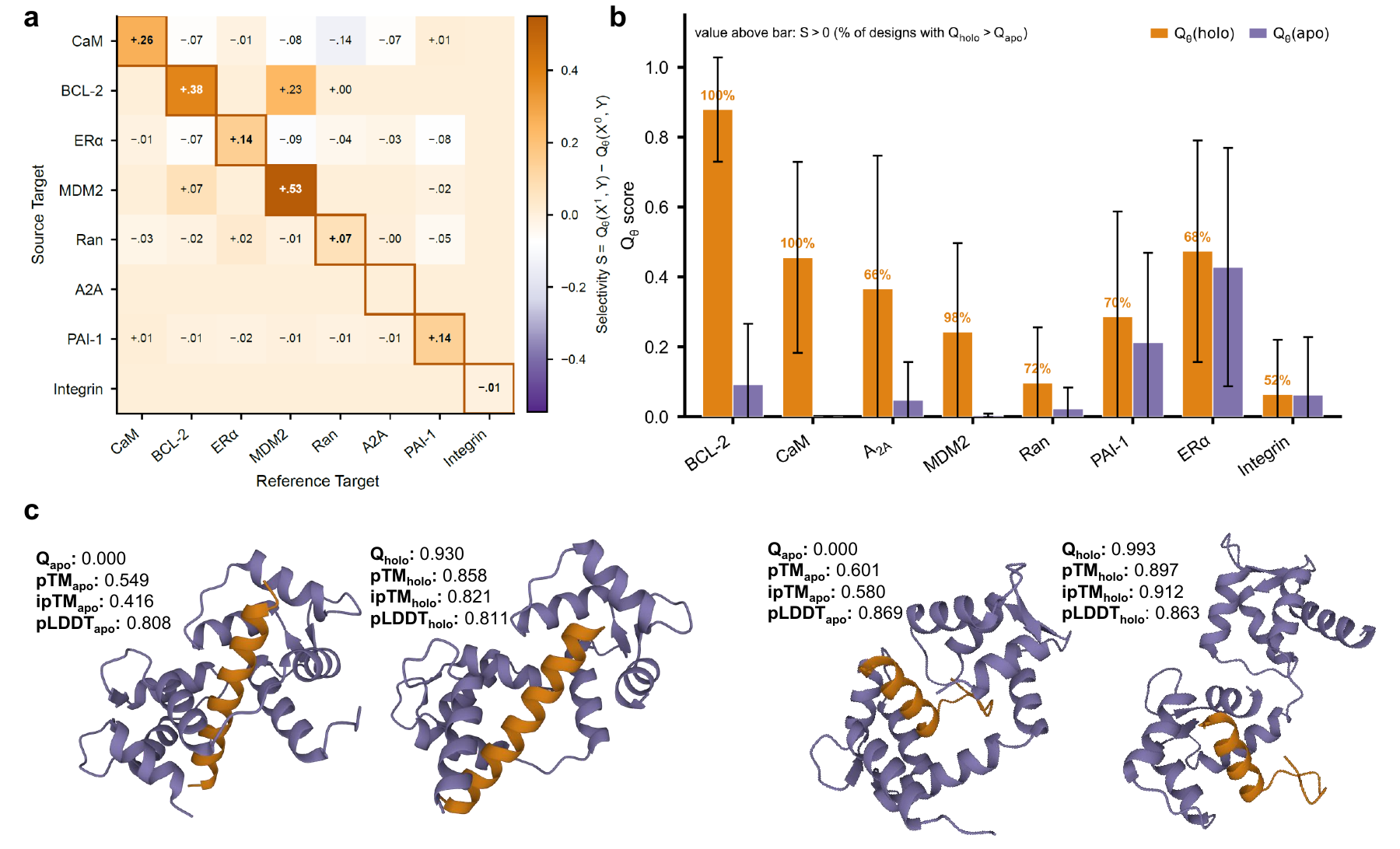}
\caption{\textbf{$Q_\theta$ conformational selectivity and CaM selectivity design.}
\textbf{(a)} Cross target selectivity matrix between Source Target (binders generated for) and the Reference Target (scored against). Zero values (0.00) are omitted. \textbf{(b)} Holo vs.\ apo $Q_\theta$ scores for 50 vanilla designs per target;
\textbf{(c)} Selectivity-based design on CaM. Two case binders (orange) shown against the apo (1st and 3rd panels) and holo (2nd and 4th panels) receptor conformations (purple).}
\label{fig:conformational_selectivity}
\end{figure*}

\subsection{Conformational Preference of \texorpdfstring{$Q_\theta$}{Q\_theta}}
\label{sec:exp_conformational}
To extend the ranking evaluation to state preference, we compared $Q_\theta$ with contact-based proxies and assessed target specificity and holo--apo score differences.

\paragraph{Comparison against energy-based baselines.}
We compared $Q_\theta$ with three contact-based proxies on the eight test targets: PRODIGY~\citep{xue2016prodigy}, interface residue count, and cross-chain edge density (Table~\ref{tab:energy_baselines}). Their mean correlations with DockQ were $+0.143$, $-0.154$, and $-0.070$, respectively, compared with $0.520$ for $Q_\theta$. PRODIGY retained positive correlation on several targets, including ER$\alpha$ ($0.551$). Across the benchmark, $Q_\theta$ ranked interface quality more accurately than these proxies.

\paragraph{Cross-target specificity.}
To test whether high scores reflected target-specific compatibility, we scored each target's 50 vanilla binders against all eight test receptors (Figure~\ref{fig:conformational_selectivity}a). The ratio of diagonal to off-diagonal scores was $19.8$, supporting target-specific compatibility. We next tested whether this compatibility differed between conformations of the same target.

\paragraph{Per-target apo rejection.}
For this within-target comparison, we scored 50 vanilla designs per target against both holo and apo structures (Figure~\ref{fig:conformational_selectivity}b). Single-seed scoring favored holo on seven of eight targets; MDM2 and BCL-2 had positive gaps for all designs (Appendix~\ref{app:apo_rejection}). The three-seed ensemble yielded $100\%$ holo preference for CaM and $98\%$ for MDM2. Integrin remained difficult, with an ensemble mean gap of $+0.001$ and $52\%$ positive gaps.
\begin{table}[htbp]
\centering
\caption{Complete target-level consensus selectivity for all 15 generator--guidance combinations, grouped by generator. Mean averages the eight targets. Bold indicates the highest value within each generator and target, or in the Mean column.}
\label{tab:results}
\setlength{\tabcolsep}{3pt}
\begin{tabular}{l*{9}{r}}
\toprule
Guidance & CaM & BCL-2 & ER$\alpha$ & A$_{2A}$ & MDM2 & PAI-1 & Ran & Integrin & Mean \\
\midrule
\multicolumn{10}{l}{\textit{RFdiffusion}} \\
Vanilla & +0.455 & +0.787 & +0.046 & +0.318 & +0.238 & +0.073 & +0.074 & +0.001 & +0.249 \\
Classifier & +0.427 & +0.741 & +0.054 & +0.332 & +0.271 & +0.054 & +0.091 & +0.006 & +0.247 \\
Langevin & \textbf{+0.677} & +0.806 & +0.106 & +0.377 & +0.366 & +0.061 & +0.108 & +0.008 & +0.314 \\
SMC & +0.510 & +0.841 & \textbf{+0.325} & +0.760 & +0.641 & \textbf{+0.389} & +0.424 & +0.013 & \textbf{+0.488} \\
TDS & +0.367 & \textbf{+0.880} & +0.050 & \textbf{+0.924} & \textbf{+0.769} & +0.262 & \textbf{+0.485} & \textbf{+0.079} & +0.477 \\
\midrule
\multicolumn{10}{l}{\textit{PXDesign}} \\
Vanilla & +0.517 & +0.560 & -0.031 & +0.023 & +0.208 & -0.000 & +0.033 & +0.015 & +0.166 \\
Classifier & +0.521 & +0.561 & -0.029 & +0.037 & +0.227 & -0.000 & +0.056 & +0.024 & +0.175 \\
Langevin & +0.022 & +0.568 & -0.027 & +0.048 & +0.262 & -0.000 & +0.084 & +0.033 & +0.124 \\
SMC & \textbf{+0.545} & +0.868 & \textbf{+0.117} & +0.377 & +0.590 & -0.000 & +0.446 & +0.027 & +0.371 \\
TDS & +0.514 & \textbf{+0.969} & +0.029 & \textbf{+0.445} & \textbf{+0.613} & -0.001 & \textbf{+0.601} & \textbf{+0.185} & \textbf{+0.419} \\
\midrule
\multicolumn{10}{l}{\textit{Proteina-ComplexA}} \\
Vanilla & +0.338 & +0.774 & -0.000 & -0.006 & +0.506 & +0.053 & +0.081 & -0.041 & +0.213 \\
Classifier & +0.432 & +0.805 & -0.000 & -0.004 & +0.567 & +0.064 & +0.128 & -0.002 & +0.249 \\
Langevin & +0.429 & +0.826 & -0.000 & -0.003 & +0.598 & +0.069 & +0.204 & -0.011 & +0.267 \\
SMC & \textbf{+0.565} & +0.836 & -0.000 & -0.000 & +0.794 & +0.123 & +0.469 & +0.017 & +0.350 \\
TDS & +0.374 & \textbf{+0.898} & +0.000 & -0.000 & \textbf{+0.883} & \textbf{+0.757} & \textbf{+0.662} & \textbf{+0.190} & \textbf{+0.470} \\
\bottomrule
\end{tabular}
\end{table}

\subsection{End-to-End Binder Design}
\label{sec:exp_design}
To test how these scoring properties translated to design, we combined three generators, RFdiffusion~\citep{watson2023novo}, PXDesign~\citep{team2025pxdesign}, and Proteina-ComplexA~\citep{didi2026scaling}, with vanilla generation and four guidance strategies: classifier guidance, SMC, TDS, and gradient ascent (Langevin) refinement. We evaluated the resulting 15 generator$\times$guidance combinations on the 8 held-out OOD test targets. Each design was scored by an ensemble of three independently trained Augmented-S2 $Q_\theta$ checkpoints.

\paragraph{Generator $\times$ guidance benchmark across OOD targets.}
Cross-target means favored TDS and SMC for each generator (Table~\ref{tab:results}). Langevin refinement improved RFdiffusion ($0.249$ to $0.314$) and Proteina-ComplexA ($0.213$ to $0.267$), but reduced PXDesign selectivity ($0.166$ to $0.124$). BCL-2 had the highest target mean ($0.781$), whereas Integrin, ER$\alpha$, and PAI-1 had lower means. ER$\alpha$ combined strong scoring correlation ($0.664$) with weak design selectivity ($0.043$), suggesting that candidate generation also constrained performance. Full target-level results are in Table~\ref{tab:results}. Candidate counts and sampling costs differed across methods (Appendix~\ref{app:design_details}). Target means across all 15 combinations were CaM $0.446$, BCL-2 $0.781$, ER$\alpha$ $0.043$, A$_{2A}$ $0.242$, MDM2 $0.502$, PAI-1 $0.127$, Ran $0.263$, and Integrin $0.038$ (grand mean $0.305$).

\begin{figure*}[htbp]
\centering
\includegraphics[width=\linewidth]{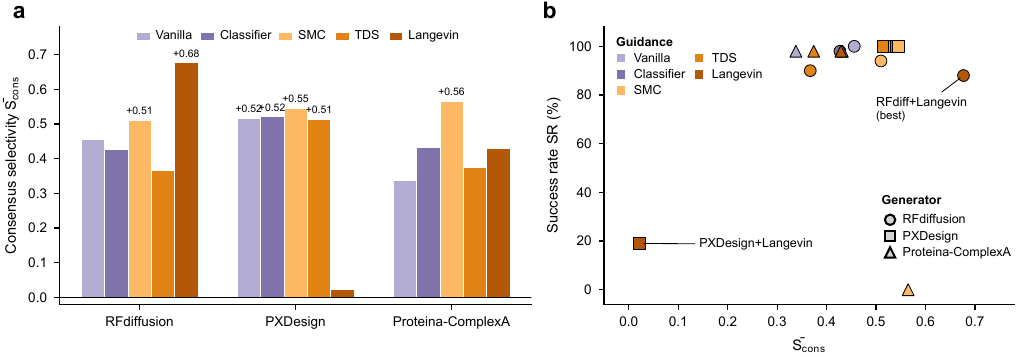}
\caption{\textbf{Generation benchmark on CaM.}
\textbf{(a)} Consensus selectivity $\bar{S}_{\mathrm{cons}}$ across 15 generation $\times$ guidance approaches. \textbf{(b)} Selectivity vs.\ design success rate (designable\,$\times$\,selective) across all generator $\times$ guidance combinations.}
\label{fig:gen_benchmark}
\end{figure*}

\paragraph{End-to-end CaM-selective design.}
For a detailed assessment of the complete workflow, we focused on CaM, whose Ca$^{2+}$-dependent rearrangement exposes a peptide-binding cleft. All 15 combinations had positive mean consensus selectivity on CaM. RFdiffusion with Langevin refinement reached $\bar{S}_{\mathrm{cons}}=0.677$ with an $88\%$ joint computational success rate, compared with $100\%$ success for vanilla RFdiffusion (Figure~\ref{fig:gen_benchmark}, Table~\ref{tab:design_main}). Gradient-noise analysis was consistent with reduced reliability of classifier guidance at high noise (Appendix~\ref{app:gradient_reliability}).

With $Q_\theta$-based reranking of a fixed vanilla CaM pool, best-of-5 and best-of-10 selection reached $\bar{S}=0.787$ and $0.885$, respectively (Table~\ref{tab:bok_sweep}). Reranking adds scoring cost to an existing pool, whereas refinement updates individual candidates. Two predicted CaM complexes had holo scores of $0.930$ and $0.993$ with near-zero apo scores (Figure~\ref{fig:conformational_selectivity}c). To assess structural changes during refinement, we checked backbone geometry across eight targets. We detected no Ramachandran outliers after refinement and measured a mean bond-length deviation of $0.005$\,\AA{} (Table~\ref{tab:structural_validity}). Changes in native-contact recovery ranged from $-0.004$ to $0.168$ on the six targets with references (Table~\ref{tab:phase3_generation}).

\paragraph{Independent computational cross-checks.}
To assess the designs beyond the optimization scorer, we compared $Q_\theta$ with independent computational metrics. Boltz-2 $\Delta$ipTM correlated with $Q_\theta$ selectivity on A$_{2A}$ ($\rho=0.500$) and CaM ($\rho=0.349$), but correlations were weak on the other targets (Table~\ref{tab:boltz2_validation}). AlphaFold~3 favored holo on the training target ALK and held-out ER$\alpha$, but favored apo on most BCL-2 designs (Table~\ref{tab:af3_iptm}). Rosetta interface energies were most favorable for BCL-2 ($-9.1$ REU) and least favorable for Integrin ($39.1$ REU) in the separate panel reported in Table~\ref{tab:rosetta_ood}. ProteinMPNN gave positive mean $\Delta$NLL on five of eight targets, and refinement reduced this metric on five targets despite increasing $Q_\theta$ (Table~\ref{tab:phase3_mpnn}). Agreement varied across these computational metrics, motivating direct binding measurements. Additional structural checks and failure analysis are in Appendix~\ref{app:design_details}.

\subsection{Experimental Validation on Calmodulin}
\label{sec:exp_wetlab}
\begin{figure}[htbp]
\centering
\includegraphics[width=\linewidth]{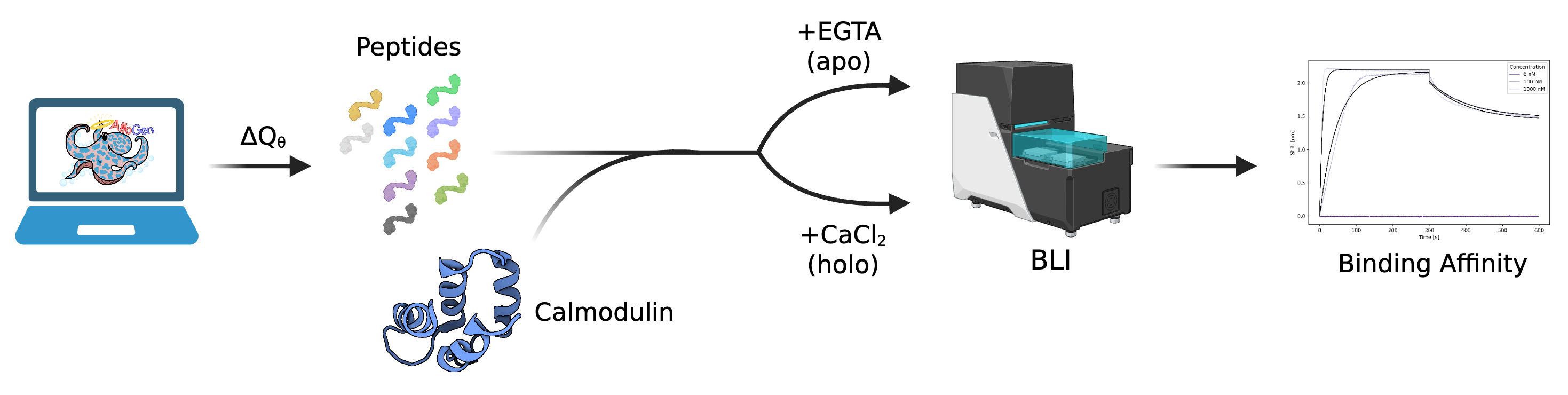}
\caption{BLI workflow for the selected CaM peptide panel. Holo and apo assay conditions used CaCl$_2$ and EGTA, respectively. Measurements were performed in duplicate and fitted with a 1:1 binding model. The schematic summarizes the assay procedure.}
\label{fig:cam_bli}
\end{figure}

To test whether selected designs bound CaM preferentially under holo conditions, we evaluated ten designed peptides, comprising nine candidates and one lower-scoring negative control, together with an M13 positive control. Candidates came from RFdiffusion and Proteina-ComplexA variants and were selected using the predicted gap $\Delta_q=Q_\theta(X^1,Y)-Q_\theta(X^0,Y)$. BLI measurements compared CaCl$_2$-supplemented holo CaM with EGTA-treated apo CaM in duplicate (Figure~\ref{fig:cam_bli}; Appendix~\ref{app:wetlab}).

Five designed peptides showed detectable holo CaM binding, including RFdiffusion designs 944 and 13 with fitted $K_D$ values of $46.6$ and $86.5$\,nM. The other positive designs had reported values of $413$\,nM, $>1.00\,\mu$M, and $1.06\,\mu$M (Table~\ref{tab:experimental_validation}). None of the designed peptides showed detectable apo CaM binding under the tested conditions. M13 bound holo CaM, and the designated negative control bound neither state. Four higher-scoring candidates also showed no detectable binding, including one with $\Delta_q=0.993$. These measurements established state-dependent binding in selected CaM designs.

\begin{table}[htbp]
\centering
\caption{BLI results for all ten designed peptides and M13. RF: RFdiffusion; Pro: Proteina-ComplexA; V: vanilla; Lg: Langevin; SM: SMC. The number is the original design or particle identifier. NB: no detectable binding under the tested conditions. M13 bound holo CaM without a reported fitted $K_D$.}
\label{tab:experimental_validation}
\begin{tabular}{llrrr}
\toprule
Generator & Design & Holo $K_D$ (nM) & Apo $K_D$ & $\Delta_q$ \\
\midrule
RF/V & 944 & 46.6 & NB & 0.413 \\
RF/V & 13 & 86.5 & NB & 0.450 \\
Pro/SM & 0273 & 413 & NB & 0.932 \\
RF/V & 657 & $>1000$ & NB & 0.930 \\
RF/Lg & 25 & 1060 & NB & 0.930 \\
Pro/V & 0145 & NB & NB & 0.993 \\
RF/Lg & 330 & NB & NB & 0.905 \\
Pro/V & 0376 & NB & NB & 0.688 \\
Pro/SM & 0954 & NB & NB & 0.912 \\
Pro/V & 0105 (negative) & NB & NB & 0.228 \\
M13 & Positive control & Bound & NB & -- \\
\bottomrule
\end{tabular}
\end{table}
\FloatBarrier
\section{Conclusion}
\label{sec:conclusion}
We introduce AlloGen to connect differential state scoring with binder reranking and guidance across frozen generators. The benchmark supports transfer to held-out targets, with guidance gains that depend on the generator and target. In selected CaM designs, BLI measurements confirmed five peptides with detectable holo binding and no detectable apo binding under the tested conditions. Extending this approach to broader target validation, richer apo decoys, and score calibration will help establish its scope for conformation-selective binder design (Appendix~\ref{sec:discussion}).

\clearpage
\begin{ack}
We thank the experimental team of the Chatterjee Laboratory for assistance with peptide design validation and helpful discussions throughout this project. We especially thank Adaptyv Bio for performing bio-layer interferometry measurements and assisting with experimental characterization of candidate binders. We thank Lauren Hong for providing the model logo.

This research was supported by a grant from the High-throughput Institute for Discovery (HIT-ID) at the University of Pennsylvania to the laboratory of Pranam Chatterjee. The work described in this paper was also partially supported by the Research Grants Council of the Hong Kong Special Administrative Region, China, under Project T45-401/22-N.

\section*{Competing Interests}
P.C. is a co-founder of Gameto, Inc., UbiquiTx, Inc., AtomBioworks, Inc., and Kosha Sciences, Inc., and advises companies involved in biologics and therapeutic development. P.C.'s interests are reviewed and managed by the University of Pennsylvania in accordance with its conflict-of-interest policies. The remaining authors declare no competing interests.
\end{ack}

\bibliography{references}
\bibliographystyle{abbrvnat}
\clearpage
\appendix

\FloatBarrier
\section{Implementation Details}
\label{app:implementation}
\FloatBarrier
\subsection{Feature Computation}

\paragraph{Backbone frames.}
Backbone frames $R_i \in \mathrm{SO}(3)$ are constructed from the $(N, C_\alpha, C)$ triplet of each residue using the Gram--Schmidt procedure. Torsion angles $(\varphi, \psi, \omega)$ are computed from four consecutive backbone atoms using the standard dihedral angle formula. At chain termini, missing torsion angles are set to zero.

\paragraph{RBF distance encoding.}
The 16-dimensional RBF encoding of distance $d$ uses Gaussian basis functions $\exp(-\gamma(d - \mu_k)^2)$ with centres $\mu_k$ evenly spaced from 2 to 12\,\AA\ and width $\gamma = 1.5$.

\FloatBarrier
\subsection{Model and Training Hyperparameters}
\label{app:hyperparams}

\paragraph{Model size.}
The InterfaceGNN scorer has $\sim\!898$K trainable parameters: 4 graph transformer layers with hidden dimension 128, 8 attention heads, query/key/value and edge bias projections, and a 3-layer scoring MLP.

\paragraph{Training hyperparameters.}
Hyperparameters for both training phases are listed in Table~\ref{tab:hyperparams}.

\begin{table}[htbp]
\centering
\caption{Training hyperparameters for the two-phase curriculum.}
\label{tab:hyperparams}
\begin{tabular}{lll}
\toprule
\textbf{Hyperparameter} & \textbf{Phase 1} & \textbf{Phase 2} \\
\midrule
Epochs          & 40               & 15               \\
Learning rate   & $5\times10^{-4}$ & $4\times10^{-5}$ \\
Weight decay    & $10^{-3}$        & $10^{-3}$        \\
Batch size      & 512              & 256              \\
Warmup steps    & 100              & --              \\
Loss            & MSE ($\mathcal{L}_q$) & InfoNCE ($\mathcal{L}_q$ dropped) \\
InfoNCE $\tau$  & --              & 0.1              \\
LR schedule     & Cosine           & Cosine           \\
Dropout         & 0.1              & 0.1              \\
Binder dropout  & 0.3              & 0.3              \\
\bottomrule
\end{tabular}
\end{table}

\paragraph{Guidance hyperparameters.}
\label{app:guidance}
For RFdiffusion classifier guidance, $Q_\theta$ is used as a guiding potential with weight $\omega{=}1.0$ and global \texttt{guide\_scale}${}=5.0$, constant decay schedule, and 200 denoising steps. For PXDesign classifier guidance, guidance scale $1.0$ is applied during diffusion progress fractions $[0.1, 0.8]$, with noise-fraction-proportional decay within that window. Langevin refinement uses step size $\eta{=}0.05$ and 100 gradient steps for RFdiffusion backbones, or $\eta{=}0.01$ and 100 steps for PXDesign. TDS uses $N{=}16$ particles and $R{=}4$ resampling rounds, retaining the top 50\% of particles each round.

\FloatBarrier
\subsection{Guidance Formulations}
\label{app:guidance_formulations}

The strategies use either gradient updates of binder backbone coordinates $\mathbf{x} = \{\mathbf{p}_i\}_{i \in Y}$ or score-based resampling. The implementations below distinguish the holo-score objective used by classifier guidance and TDS from the two-state margin used by refinement and SMC.

\paragraph{Langevin backbone refinement.}\citep{welling2011bayesian,song2021scorebased}
Given a generated backbone $\mathbf{x}_0$, we perform $T$ steps of gradient ascent on the selectivity margin:
\begin{equation}
  \mathbf{x}_{t+1} = \mathbf{x}_t + \eta\, \nabla_{\mathbf{x}} S_\theta(\mathbf{x}_t; X^1, \{X^0\}), \quad t = 0, \ldots, T{-}1,
\end{equation}
where $\eta$ is the step size. This operates on fully denoised backbones, avoiding the noisy-gradient regime.

\paragraph{Classifier guidance.}\citep{dhariwal2021diffusion}
During diffusion denoising, we inject the $Q_\theta$ gradient into each denoising step:
\begin{equation}
  \hat{\boldsymbol{\epsilon}}_t = \boldsymbol{\epsilon}_\psi(\mathbf{x}_t, t) - \omega\, \sigma_t\, \nabla_{\mathbf{x}_t} \log Q_\theta(X^1, \mathbf{x}_t),
\end{equation}
where $\boldsymbol{\epsilon}_\psi$ is the denoiser, $\omega$ the guidance weight, and $\sigma_t$ the noise level at timestep $t$.

\paragraph{Twisted diffusion sampling (TDS).}\citep{wu2023practical}
TDS uses $N$ particles $\{\mathbf{x}_t^{(n)}\}_{n=1}^N$ with importance weights $w_t^{(n)} \propto Q_\theta(X^1, \hat{\mathbf{x}}_0^{(n)})$ computed from the denoised prediction at each timestep. Multinomial resampling is applied when the effective sample size $\mathrm{ESS} = (\sum_n w_n)^2 / \sum_n w_n^2$ drops below $N/2$.

\paragraph{Sequential Monte Carlo (SMC).}\citep{wu2023practical}
SMC extends TDS with multiple resampling rounds: after generating $N$ complete trajectories, the top fraction by $S_\theta$ seeds the next round, iteratively enriching the candidate pool toward high-selectivity designs.

\FloatBarrier
\section{Dataset Construction}
\label{app:dataset}

The dataset comprises 2{,}896 complexes across 65 two-state protein targets in 15 families. Each complex yields 12 training samples: 1 native holo labeled 1.0, 1 apo mismatch labeled 0.0, and 10 rigid-body decoys with labels $\max(0, 1 - d_{\mathrm{RMSD}}/4)$, plus 959 FastRelax-based hard negatives with mean label 0.304. Interface graphs contain $34.7 \pm 19.7$ nodes, ranging from 3 to 128 with a median of 29, comprising $18.1 \pm 9.8$ receptor and $16.6 \pm 11.0$ binder residues. The target split assigns 51 targets to training, 6 to OOD validation, and 8 to OOD test with CaM held out, yielding 30{,}623 samples split as 23{,}640 training, 4{,}020 validation, and 2{,}963 test (Table~\ref{tab:split_v5}). Linked targets such as SRC and SRC-SH2 are forced into the same partition. All scoring experiments use the S2-augmented configuration totalling 39{,}867 training samples; ablations across S1/S2/S3 are detailed in Table~\ref{tab:dataset_versions}.

\begin{table}[htbp]
\centering
\caption{Dataset configurations and augmentation inventory. The target partition is 51/6/8 for training/validation/test in both configurations. The augmentation sources and configuration totals are listed separately.}
\label{tab:dataset_versions}
\begin{tabular}{lrr}
\toprule
Source & Baseline & Augmented \\
\midrule
FastRelax negatives & 959 & 959 \\
Generator decoys & -- & 4,862 \\
Cross-family negatives & -- & 254 \\
Conformational decoys & -- & 1,075 \\
\bottomrule
\end{tabular}
\par\medskip
\begin{tabularx}{\linewidth}{lXr}
\toprule
Setting & Composition & Samples \\
\midrule
S1 (Base) & Native complexes, rigid-body decoys, and FastRelax negatives & 34,751 \\
S2 (Full) & S1 plus generator, cross-family, and conformational decoys & 39,867 \\
S3 (Partial) & S1 plus cross-family and conformational decoys & 36,080 \\
\bottomrule
\end{tabularx}
\end{table}

\begin{table}[htbp]
\centering
\caption{Training targets by family (51 total). CaM belongs to the test partition and ALK to training. Validation and test target identities are listed in Table~\ref{tab:heldout_targets}.}
\label{tab:split_v5}\label{tab:expanded_families}
\begin{tabularx}{\linewidth}{lX}
\toprule
Family (total targets) & Training targets \\
\midrule
Kinase (9) & ABL1, \textbf{ALK}, Aurora~A, BRAF, CDK2, EGFR, GRK2, SRC, SRC-SH2 \\
Small GTPase (6) & Arf1, CDC42, KRAS, RhoA \\
Nuclear Receptor (5) & AR, GR, PPAR$\gamma$, RXR$\alpha$ \\
GPCR/Ion Ch.\ (6) & CFTR, KcsA, nAChR \\
Phosphatase (4) & Calcineurin, PP2A, PTP1B, SHP2 \\
Protease (6) & HIV-PR, MMP-2, M$^{\text{pro}}$, Thrombin, USP7 \\
BCL-2 (2) & BCL-xL \\
14-3-3 (2) & -- \\
Ca sensor (3) & S100B, Troponin~C \\
G protein (2) & CheY, G$\alpha_s$ \\
Chaperone/Enzyme (6) & ATCase, CypA, DHFR, Enolase, Hsp70, RNase~A \\
Epigenetic (3) & BRD4, WDR5 \\
Trafficking (5) & ABCB1, Importin-$\beta$, Maltose~BP, p97, VHL \\
Cell adhesion (3) & CTLA-4, Integrin~$\beta_3$ \\
Signaling (3) & Gelsolin, Hemoglobin \\
\bottomrule
\end{tabularx}
\end{table}

\begin{table}[htbp]
\centering
\caption{Held-out validation (6) and test (8) targets for the target-level split. The test partition contains CaM and excludes ALK.}
\label{tab:heldout_targets}
\begin{tabularx}{\linewidth}{lXX}
\toprule
Family & Validation & Test \\
\midrule
Small GTPase (6) & Rac1 & Ran \\
Nuclear Receptor (5) & -- & ER$\alpha$ \\
GPCR/Ion Ch.\ (6) & $\beta_2$AR, $\mu$-opioid & A$_{2A}$ \\
Protease (6) & Caspase-3 & -- \\
BCL-2 (2) & -- & BCL-2 \\
14-3-3 (2) & 14-3-3, 14-3-3$\sigma$ & -- \\
Ca sensor (3) & -- & \textbf{CaM} \\
Epigenetic (3) & -- & MDM2 \\
Cell adhesion (3) & -- & Integrin \\
Signaling (3) & -- & PAI-1 \\
\bottomrule
\end{tabularx}
\end{table}

\FloatBarrier
\section{Extended \texorpdfstring{$Q_\theta$}{Q\_theta} Results}
\label{app:scorer_details}

All results in this section use the S2 configuration with the target split (Table~\ref{tab:split_v5}), ESM-2 encoder, and binder-dropout rate $p{=}0.3$, consistent with the experimental setup in Section~\ref{sec:exp_scoring}. The suffix S1/S2/S3 denotes the dataset augmentation configuration as defined in Table~\ref{tab:dataset_versions}.

\FloatBarrier
\subsection{Per-Target Selectivity Performance}
Phase~2 improved mean scoring correlation from $0.481$ to $0.520$ (Table~\ref{tab:phase_ablation}). Gains were largest on Integrin ($0.166$) and A$_{2A}$ ($0.170$), followed by MDM2 ($0.037$) and PAI-1 ($0.027$). Correlation decreased on BCL-2 ($-0.049$), ER$\alpha$ ($-0.025$), and CaM ($-0.017$), and was unchanged on Ran. The mean gain therefore reflected improvements on some targets and decreases on others.
\begin{table}[htbp]
\centering
\caption{Phase~1 vs.\ Phase~1+2 Spearman $\rho$ on 8 OOD targets under the target split. Phase~2 values are 3-seed means $\pm$ std.}
\label{tab:phase_ablation}
\begin{tabular}{lcc}
\toprule
\textbf{Target} & Phase~1 $\rho$ & Phase~2 $\rho$ \\
\midrule
BCL-2      & $0.716$ & $0.667 \pm 0.014$ \\
ER$\alpha$ & $0.689$ & $0.664 \pm 0.013$ \\
MDM2       & $0.552$ & $0.589 \pm 0.036$ \\
CaM        & $0.491$ & $0.474 \pm 0.061$ \\
Ran        & $0.511$ & $0.511 \pm 0.065$ \\
A$_{2A}$   & $0.299$ & $0.469 \pm 0.034$ \\
PAI-1      & $0.394$ & $0.421 \pm 0.045$ \\
Integrin   & $0.200$ & $0.366 \pm 0.085$ \\
\midrule
\textbf{Mean (8)} & $0.481$ & $\mathbf{0.520} \pm 0.010$ \\
\bottomrule
\end{tabular}
\end{table}

\FloatBarrier
\subsection{Phase 2 Batch-Size Ablation}
\label{app:phase2_ablation}
To assess sensitivity to the number of contrastive negatives, we varied the Phase~2 batch size; all three InfoNCE settings exceeded the Phase~1-only result. Batch size 256 had the highest mean correlation ($0.530$), followed by 512 ($0.500$) and 64 ($0.496$) (Table~\ref{tab:phase2_ablation}). Both negative diversity and optimization behavior varied with batch size, leaving their individual contributions unresolved.

\begin{table}[htbp]
\centering
\caption{Phase~2 ablation across InfoNCE batch sizes. All variants start from the same Phase~1 checkpoint.}\label{tab:phase2_ablation}

\begin{tabular}{lc}
\toprule
\textbf{Configuration} & $\bar\rho$ \\
\midrule
\multicolumn{2}{l}{\emph{Paired InfoNCE (cross-target negatives)}} \\
InfoNCE, BS$=$256   & $\mathbf{0.530}$ \\
InfoNCE, BS$=$512   & $0.500$ \\
InfoNCE, BS$=$64    & $0.496$ \\
\midrule
Phase~1 only (MSE)  & $0.489$ \\
\bottomrule
\end{tabular}
\end{table}

\FloatBarrier
\subsection{Scores Along Interpolated Conformations}
\label{app:conformational_path}
To examine score variation between the endpoint states, we scored CaM binders against 11 interpolated receptor structures between apo and holo ($\tau=0$ to $1$). All ten vanilla designs had positive Spearman correlation between interpolation position and score, with mean $\rho=0.518$ (Table~\ref{tab:conf_landscape}). Langevin-refined designs had earlier score onset ($\tau_{50\%}=0.4$--$0.9$, compared with $0.8$--$1.0$). These correlations describe overall score trends along geometric interpolations; strict monotonicity and physical transition paths were not established.

\begin{table}[htbp]
\centering
\caption{CaM interpolation analysis: 20 refined and 10 vanilla designs, each scored on 11 frames. The positive fraction refers to Spearman correlation with interpolation position, not strict monotonicity.}
\label{tab:conf_landscape}
\begin{tabular}{lrrrr}
\toprule
Set & Mean $\rho$ & $\rho>0$ & Best $\rho$ & $\tau_{50\%}$ \\
\midrule
Vanilla (10) & $\mathbf{+0.518}$ & $100\%$ (10/10) & $0.745$ & $0.8$--$1.0$ \\
Langevin (20) & $+0.246$ & $80\%$ (16/20) & $0.891$ & $0.4$--$0.9$ \\
\bottomrule
\end{tabular}
\end{table}

\FloatBarrier
\subsection{Apo Rejection Analysis}
\label{app:apo_rejection}
Under single-seed scoring, seven of eight targets had positive mean holo--apo gaps, and all MDM2 and BCL-2 designs had positive gaps (Table~\ref{tab:apo_rejection}). Integrin had a marginal mean apo preference ($\Delta Q=-0.009$) and $40\%$ positive gaps. Under the three-seed ensemble, CaM had mean holo and apo scores of $0.456$ and $0.000$, with $100\%$ positive gaps.

\begin{table}[htbp]
\centering
\caption{Holo vs.\ apo $Q_\theta$ scores for 50 vanilla designs per target.}
\label{tab:apo_rejection}
\begin{tabular}{lcccc}
\toprule
\textbf{Target} & $\bar{Q}_{\text{holo}}$ & $\bar{Q}_{\text{apo}}$ & $\Delta Q$ & $S{>}0$ \\
\midrule
MDM2       & $0.577 \pm 0.189$ & $0.003 \pm 0.006$ & $+0.575$ & $100\%$ \\
BCL-2      & $0.485 \pm 0.281$ & $0.000 \pm 0.000$ & $+0.485$ & $100\%$ \\
ER$\alpha$ & $0.342 \pm 0.250$ & $0.120 \pm 0.053$ & $+0.222$ & $76\%$  \\
CaM        & $0.213 \pm 0.136$ & $0.065 \pm 0.052$ & $+0.148$ & $74\%$  \\
PAI-1      & $0.146 \pm 0.093$ & $0.018 \pm 0.057$ & $+0.128$ & $90\%$  \\
Ran        & $0.065 \pm 0.057$ & $0.000 \pm 0.002$ & $+0.064$ & $92\%$  \\
A$_{2A}$   & $0.062 \pm 0.096$ & $0.000 \pm 0.000$ & $+0.062$ & $92\%$  \\
Integrin   & $0.064 \pm 0.054$ & $0.072 \pm 0.058$ & $-0.009$ & $40\%$  \\
\bottomrule
\end{tabular}
\end{table}

\FloatBarrier
\subsection{Multi-Seed Scoring Robustness}

In the additional CaM scoring comparison, vanilla and Langevin gaps varied little across three training seeds (Table~\ref{tab:multiseed}). These seed identifiers and scores are reported separately from the consensus generation benchmark in Table~\ref{tab:design_main}.

\begin{table}[htbp]
\centering
\caption{Per-seed $Q_\theta$ selectivity $\bar{S}$ on CaM under three independently trained seeds.}
\label{tab:multiseed}
\begin{tabular}{llccccc}
\toprule
& & \multicolumn{3}{c}{\textbf{Per-seed $\bar{S}$}} \\
\cmidrule(lr){3-5}
\textbf{Target} & \textbf{Guidance} & s2048 & s10040 & s10043 & Mean & $\sigma$ \\
\midrule
\multirow{2}{*}{CaM}
  & Vanilla  & $+0.148$ & $+0.146$ & $+0.156$ & $+0.150$ & $0.004$ \\
  & Langevin & $+0.255$ & $+0.250$ & $+0.263$ & $\mathbf{+0.256}$ & $0.005$ \\
\bottomrule
\end{tabular}
\end{table}

\FloatBarrier
\section{Extended Design Results}
\label{app:design_details}

\FloatBarrier
\subsection{Generation Benchmark Performance}

We evaluated CaM design using RFdiffusion~\citep{watson2023novo}, PXDesign~\citep{team2025pxdesign}, and Proteina-ComplexA~\citep{didi2026scaling}, each with vanilla sampling and four guidance strategies. Selectivity was evaluated with three independently trained scorer checkpoints (Table~\ref{tab:design_main}); structural and sequence metrics are listed in Table~\ref{tab:cam_structure}.

\begin{table}[htbp]
\centering
\caption{CaM design selectivity and computational success rates. RF: RFdiffusion; PX: PXDesign; Pro: Proteina-ComplexA. All $S$ values are bounded score gaps. Des: designable fraction; SR: reported fraction satisfying designability and positive selectivity. Structural metrics are listed in Table~\ref{tab:cam_structure}.}
\label{tab:design_main}
\begin{tabular}{lrrrrrr}
\toprule
Method & $N$ & $\bar S_{\rm cons}$ & Top-5 $\bar S$ & $S>0$ & Des & SR \\
\midrule
RF/Langevin & 50 & $\mathbf{+0.677}$ & $\mathbf{0.993}$ & $100\%$ & $88\%$ & $\mathbf{88\%}$ \\
RF/SMC & 64 & $+0.510$ & $0.941$ & $100\%$ & $94\%$ & $94\%$ \\
RF/Vanilla & 50 & $+0.456$ & $0.932$ & $100\%$ & $100\%$ & $100\%$ \\
RF/Classifier & 50 & $+0.427$ & $0.891$ & $100\%$ & $98\%$ & $98\%$ \\
RF/TDS & 40 & $+0.367$ & $0.818$ & $100\%$ & $90\%$ & $90\%$ \\
PX/SMC & 64 & $+0.545$ & $0.851$ & $100\%$ & $100\%$ & $100\%$ \\
PX/Classifier & 50 & $+0.521$ & $0.891$ & $100\%$ & $100\%$ & $100\%$ \\
PX/Vanilla & 50 & $+0.517$ & $0.870$ & $100\%$ & $100\%$ & $100\%$ \\
PX/TDS & 64 & $+0.514$ & $0.851$ & $100\%$ & $100\%$ & $100\%$ \\
PX/Langevin & 50 & $+0.022$ & $0.424$ & $24\%$ & $80\%$ & $19\%$ \\
Pro/SMC & 52 & $\mathbf{+0.565}$ & $\mathbf{0.992}$ & $\mathbf{100\%}$ & $0\%$ & $0\%$ \\
Pro/Classifier & 52 & $+0.432$ & $0.975$ & $98\%$ & $100\%$ & $98\%$ \\
Pro/Langevin & 52 & $+0.429$ & $0.975$ & $98\%$ & $100\%$ & $98\%$ \\
Pro/TDS & 52 & $+0.374$ & $0.981$ & $98\%$ & $100\%$ & $98\%$ \\
Pro/Vanilla & 52 & $+0.338$ & $0.942$ & $98\%$ & $100\%$ & $98\%$ \\
\bottomrule
\end{tabular}
\end{table}
\begin{table}[htbp]
\centering
\caption{Structural and sequence metrics for the CaM designs in Table~\ref{tab:design_main}. ipTM is from Boltz-2; scTM is structural self-consistency; NLL is the ProteinMPNN holo-state negative log-likelihood. Proteina-ComplexA generates full-atom complexes with co-designed sequences.}
\label{tab:cam_structure}
\begin{tabular}{lrrr}
\toprule
Method & ipTM & scTM & NLL$_{\rm holo}$ \\
\midrule
RF/Langevin & $0.742$ & $0.806$ & $\mathbf{1.39}$ \\
RF/SMC & $0.733$ & $0.828$ & $3.60$ \\
RF/Vanilla & $\mathbf{0.769}$ & $0.820$ & $3.59$ \\
RF/Classifier & $0.715$ & $0.826$ & $3.54$ \\
RF/TDS & $0.766$ & $\mathbf{0.835}$ & $3.66$ \\
PX/SMC & $0.710$ & $0.698$ & $3.71$ \\
PX/Classifier & $0.757$ & $0.684$ & $3.74$ \\
PX/Vanilla & $0.736$ & $0.690$ & $3.74$ \\
PX/TDS & $0.762$ & $0.690$ & $3.74$ \\
PX/Langevin & $0.639$ & $0.664$ & $1.58$ \\
Pro/SMC & $0.420$ & $0.340$ & $1.38$ \\
Pro/Classifier & $0.399$ & $0.597$ & $1.88$ \\
Pro/Langevin & $0.381$ & $0.596$ & $1.88$ \\
Pro/TDS & $0.376$ & $0.592$ & $1.86$ \\
Pro/Vanilla & $0.387$ & $0.601$ & $1.80$ \\
\bottomrule
\end{tabular}
\end{table}

\FloatBarrier
\subsection{Structural Validity of Langevin Refinement}
\label{app:structural_validity}
To check whether refinement altered local backbone geometry, we compared 50 designs per target before and after refinement (Table~\ref{tab:structural_validity}). Both sets had zero detected Ramachandran outliers across eight targets. Mean $\omega$ deviation decreased from $4.2^\circ$ to $2.3^\circ$, while mean bond-length deviation increased from $0.000$ to $0.005$\,\AA. These measurements characterize local geometry; global stability and foldability require separate assessment.

\begin{table}[htbp]
\centering
\caption{Backbone geometry for vanilla and Langevin-refined designs across 8 OOD targets. No Ramachandran outliers were detected in either set.}
\label{tab:structural_validity}
\begin{tabular}{lcccccc}
\toprule
& \multicolumn{2}{c}{Rama outlier (\%)} & \multicolumn{2}{c}{$\omega$ dev ($^\circ$)} & \multicolumn{2}{c}{Bond dev (\AA)} \\
\cmidrule(lr){2-3} \cmidrule(lr){4-5} \cmidrule(lr){6-7}
\textbf{Target} & Van & Lang & Van & Lang & Van & Lang \\
\midrule
CaM        & $0.0$ & $0.0$ & $4.0$ & $4.5$ & $0.000$ & $0.014$ \\
ER$\alpha$ & $0.0$ & $0.0$ & $4.5$ & $2.0$ & $0.000$ & $0.006$ \\
BCL-2      & $0.0$ & $0.0$ & $4.3$ & $1.4$ & $0.000$ & $0.001$ \\
MDM2       & $0.0$ & $0.0$ & $4.1$ & $2.3$ & $0.000$ & $0.008$ \\
Ran        & $0.0$ & $0.0$ & $3.6$ & $1.9$ & $0.000$ & $0.003$ \\
A$_{2A}$   & $0.0$ & $0.0$ & $3.9$ & $1.9$ & $0.000$ & $0.004$ \\
PAI-1      & $0.0$ & $0.0$ & $4.7$ & $2.0$ & $0.000$ & $0.003$ \\
Integrin   & $0.0$ & $0.0$ & $4.6$ & $2.6$ & $0.000$ & $0.002$ \\
\midrule
\textbf{Mean} & $0.0$ & $0.0$ & $4.2$ & $2.3$ & $0.000$ & $0.005$ \\
\bottomrule
\end{tabular}
\end{table}

\FloatBarrier
\subsection{Reranking vs.\ Gradient Refinement}
\label{app:reranking_comparison}
We compared reranking and refinement on a CaM pool with mean vanilla gap $0.456$ and $100\%$ positive gaps (Table~\ref{tab:rerank_vs_langevin}). Best-of-5 and best-of-10 selection reached $0.787$ and $0.885$, respectively, compared with $0.677$ for Langevin refinement. All gaps were evaluated with the same $Q_\theta$ scorer used for selection or optimization.

\begin{table}[htbp]
\centering
\caption{Best-of-$K$ reranking vs.\ Langevin refinement on 50 CaM vanilla designs under ensemble scoring.}
\label{tab:rerank_vs_langevin}
\begin{tabular}{lcc}
\toprule
\textbf{Strategy} & $\bar{S}$ & $S{>}0$ \\
\midrule
Vanilla (all 50)     & $+0.456$ & $100\%$ \\
Rerank best-of-2     & $+0.616$ & $100\%$ \\
Rerank best-of-5     & $+0.787$ & $100\%$ \\
Rerank best-of-10    & $+0.885$ & $100\%$ \\
Rerank best-of-20    & $+0.947$ & $100\%$ \\
Langevin (100 steps) & $\mathbf{+0.677}$ & $\mathbf{100\%}$ \\
\bottomrule
\end{tabular}
\end{table}

\FloatBarrier
\subsection{Gradient Reliability Under Noise}
\label{app:gradient_reliability}
To assess why guidance during denoising differed from refinement, we measured the cosine similarity between $Q_\theta$ gradients computed on clean and noise-perturbed backbones as a function of Gaussian noise scale $\sigma$. Cosine similarity dropped from $0.75$ at $\sigma{=}0.1\,\text{\AA}$ to $0.12$ at $\sigma{=}0.5\,\text{\AA}$, and approached zero for $\sigma{\geq}2.0\,\text{\AA}$. Gradient reliability therefore degrades rapidly with noise, consistent with the advantage of refinement on denoised backbones over guidance at higher noise levels.

\begin{table}[htbp]
\centering
\caption{$Q_\theta$ gradient cosine similarity under Gaussian backbone noise, averaged over 20 CaM designs.}
\label{tab:grad_noise}
\begin{tabular}{lccc}
\toprule
$\sigma$ (\AA) & Norm & Cos & Margin \\
\midrule
$0.0$ & $0.317$ & $\mathbf{1.000}$ & $+0.359$ \\
$0.1$ & $0.329$ & $0.753$ & $+0.352$ \\
$0.5$ & $0.368$ & $0.121$ & $+0.380$ \\
$1.0$ & $0.366$ & $0.032$ & $+0.349$ \\
$2.0$ & $0.172$ & $0.012$ & $+0.211$ \\
$5.0$ & $0.018$ & $0.009$ & $+0.015$ \\
\bottomrule
\end{tabular}
\end{table}

\FloatBarrier
\subsection{Noise Schedule Alignment}
\label{app:noise_alignment}
To relate this noise sensitivity to denoising, we aligned the RFdiffusion noise schedule ($T{=}50$ denoising steps) with the gradient-reliability profile from Appendix~\ref{app:gradient_reliability}. Cosine similarity was $0.75$ at $t{=}0$ and fell to $0.31$ by $t{=}5$ and to $0.06$ by $t{=}25$, remaining above $0.5$ only for roughly the final $2$ of $50$ steps, once the trajectory enters the regime $\sigma{<}0.10\,\text{\AA}$. About $96\%$ of steps in this 50-step diagnostic fell below a cosine similarity of $0.5$; reliability under the main 200-step guidance configuration remains to be quantified.

\begin{table}[htbp]
\centering
\caption{RFdiffusion noise schedule mapped to $Q_\theta$ gradient reliability across timesteps.}
\label{tab:noise_schedule}
\begin{tabular}{cccc}
\toprule
Timestep & $\sigma$ (\AA) & $\bar\alpha$ & Cosine \\
\midrule
$0$  & $0.10$ & $0.99$ & $0.75$ \\
$5$  & $0.28$ & $0.92$ & $0.31$ \\
$10$ & $0.40$ & $0.84$ & $0.15$ \\
$25$ & $0.70$ & $0.51$ & $0.06$ \\
$49$ & $0.93$ & $0.13$ & $0.05$ \\
\bottomrule
\end{tabular}
\end{table}

\FloatBarrier
\subsection{Langevin Step-Size Sensitivity}
\label{app:eta_sensitivity}
To assess the balance between selectivity gains and backbone displacement, we swept the Langevin step size $\eta$ on 50 CaM designs under ensemble scoring (Table~\ref{tab:eta_sweep}). Selectivity increased from $\bar{S}{=}{+}0.506$ at $\eta{=}0.01$ to $\bar{S}{=}{+}0.579$ at $\eta{=}0.08$, then plateaued, while C$\alpha$ RMSD from the vanilla backbone stayed below $0.14\,\text{\AA}$ throughout. At $\eta{=}0.04$, mean selectivity was $0.559$ and RMSD was $0.10\,\text{\AA}$. Generator-specific benchmark settings are listed in Appendix~\ref{app:guidance}.

\begin{table}[htbp]
\centering
\caption{Langevin step-size sweep on CaM under ensemble scoring. RMSD denotes C$\alpha$ displacement from the vanilla backbone.}
\label{tab:eta_sweep}
\begin{tabular}{lcccc}
\toprule
$\eta$ & $\bar{S}$ & $\Delta\bar{S}$ & $S{>}0$ & RMSD (\AA) \\
\midrule
-- (vanilla) & $+0.433$ & --      & $98\%$  & --  \\
$0.01$ & $+0.506$ & $+0.073$ & $100\%$ & $0.04$ \\
$0.02$ & $+0.533$ & $+0.100$ & $100\%$ & $0.06$ \\
$0.04$ & $+0.559$ & $+0.126$ & $100\%$ & $0.10$ \\
$0.08$ & $\mathbf{+0.579}$ & $\mathbf{+0.146}$ & $\mathbf{100\%}$ & $0.12$ \\
$0.10$ & $+0.579$ & $+0.146$ & $100\%$ & $0.14$ \\
\bottomrule
\end{tabular}
\end{table}

\FloatBarrier
\subsection{Best-of-\texorpdfstring{$K$}{K} Reranking}
\label{app:bok}
To quantify the effect of candidate-pool size under the selection scorer, we evaluated best-of-$K$ selection from a fixed CaM pool using 10{,}000 bootstrap trials (Table~\ref{tab:bok_sweep}). At $K=5$, the mean selected gap was $0.787$ and $94\%$ of selections exceeded $0.5$. At $K=10$, these values reached $0.885$ and $100\%$.

\begin{table}[htbp]
\centering
\caption{Best-of-$K$ reranking statistics on CaM with 10{,}000 bootstrap trials. ``High'' denotes $S{>}0.5$.}
\label{tab:bok_sweep}
\begin{tabular}{lccc}
\toprule
$K$ & Mean & $S{>}0$ & High \\
\midrule
$1$  & $+0.452$ & $100\%$ & $41\%$ \\
$2$  & $+0.616$ & $100\%$ & $67\%$ \\
$5$  & $+0.787$ & $100\%$ & $94\%$ \\
$10$ & $+0.885$ & $100\%$ & $100\%$ \\
$25$ & $+0.961$ & $100\%$ & $100\%$ \\
$50$ & $\mathbf{+0.985}$ & $\mathbf{100\%}$ & $\mathbf{100\%}$ \\
\bottomrule
\end{tabular}
\end{table}

\FloatBarrier
\subsection{End-to-end Langevin Design Results}
\label{app:phase3_generation}

\paragraph{Structure-related metric analysis for Langevin refinement.}
We refined 50 vanilla designs per target for 100 steps (Table~\ref{tab:phase3_generation}). Mean selectivity increased on seven of eight targets, with the largest gains on CaM ($0.253$) and MDM2 ($0.181$), and decreased on PAI-1 ($-0.024$). The fraction of retained designs with positive gaps was $100\%$ for CaM and BCL-2 and at least $70\%$ for seven targets. Across the eight targets, 390 of 400 designs were retained. Changes in native-contact recovery ranged from $-0.004$ to $0.168$ on the six targets with contact references (Table~\ref{tab:fnat}).
\begin{table}[htbp]
\centering
\caption{Refinement outcomes across the eight held-out targets. Positive-gap rate denotes the fraction of retained refined designs with $S>0$, not the fraction improved over vanilla. Of 400 original designs, 390 were retained.}
\label{tab:phase3_generation}
\begin{tabular}{lrrrrrr}
\toprule
Target & $N_{\rm van}$ & $N_{\rm ref}$ & $S>0$ & $\bar S_{\rm van}$ & $\bar S_{\rm ref}$ & $\Delta\bar S$ \\
\midrule
CaM & 50 & 50 & $100\%$ & 0.531 & 0.784 & $+0.253$ \\
ER$\alpha$ & 50 & 50 & $74\%$ & 0.110 & 0.250 & $+0.140$ \\
BCL-2 & 50 & 50 & $100\%$ & 0.886 & 0.911 & $+0.024$ \\
MDM2 & 50 & 50 & $98\%$ & 0.216 & 0.397 & $+0.181$ \\
Ran & 50 & 46 & $80\%$ & 0.119 & 0.249 & $+0.130$ \\
PAI-1 & 50 & 49 & $71\%$ & 0.188 & 0.164 & $-0.024$ \\
Integrin & 50 & 48 & $58\%$ & 0.004 & 0.012 & $+0.008$ \\
A$_{2A}$ & 50 & 47 & $70\%$ & 0.363 & 0.412 & $+0.049$ \\
\bottomrule
\end{tabular}
\end{table}
\begin{table}[htbp]
\centering
\caption{Change in native-contact recovery after refinement, measured against the available crystal-contact reference. CaM and A$_{2A}$ lack the short-peptide contact references used in this comparison.}
\label{tab:fnat}
\begin{tabular}{lr}
\toprule
Target & $\Delta$ fNAT \\
\midrule
CaM & -- \\
ER$\alpha$ & $-0.004$ \\
BCL-2 & $+0.168$ \\
MDM2 & $+0.084$ \\
Ran & $+0.002$ \\
PAI-1 & $-0.001$ \\
Integrin & $+0.012$ \\
A$_{2A}$ & -- \\
\bottomrule
\end{tabular}
\end{table}

\paragraph{ProteinMPNN $\Delta$NLL analysis for Langevin refinement.}
To test whether refinement also improved sequence-based state preference, we evaluated ProteinMPNN $\Delta$NLL. Its mean was positive for five of eight vanilla target panels, with an all-target mean of $0.188$ (Table~\ref{tab:phase3_mpnn}). Langevin refinement reduced $\Delta$NLL on five targets and reduced the all-target mean to $0.128$. Thus, higher $Q_\theta$ scores did not imply higher sequence-recovery preference for holo. The paired tests assessed changes between vanilla and refined designs in this computational measure of state preference.
\begin{table}[htbp]
\centering
\caption{ProteinMPNN $\Delta$NLL for vanilla and Langevin-refined designs. ProteinMPNN measures sequence-recovery likelihood given a backbone, a $Q_\theta$-independent metric that probes a different dimension of binder--receptor compatibility than the geometric selectivity signal $Q_\theta$ optimizes. Positive $\Delta$NLL indicates holo preference; $p$-values are from paired Wilcoxon signed-rank tests on per-design (vanilla, Langevin) pairs. $^\dag$Targets with negative mean vanilla $\Delta$NLL. The cause of the disagreement with $Q_\theta$ is unresolved.}
\label{tab:phase3_mpnn}
\begin{tabular}{lcccc}
\toprule
\textbf{Target} & $\Delta$NLL$_{\text{van}}$ & $\Delta$NLL$_{\text{lang}}$ & $\Delta$ & $p$ \\
\midrule
MDM2          & $+1.050$ & $+0.743$ & $-0.307$ & $<\!10^{-14*}$ \\
BCL-2         & $+0.550$ & $+0.436$ & $-0.114$ & $<\!10^{-14*}$ \\
CaM           & $+0.290$ & $+0.188$ & $-0.102$ & $<\!10^{-14*}$ \\
PAI-1         & $+0.120$ & $+0.095$ & $-0.026$ & $8\!\times\!10^{-6\,*}$ \\
Integrin      & $+0.014$ & $+0.016$ & $+0.002$ & $0.020^*$ \\
\midrule
ER$\alpha^\dag$ & $-0.129$ & $-0.054$ & $+0.075$ & $<\!10^{-12*}$ \\
Ran$^\dag$      & $-0.169$ & $-0.136$ & $+0.033$ & $6\!\times\!10^{-5\,*}$ \\
A$_{2A}^\dag$   & $-0.223$ & $-0.261$ & $-0.040$ & $<\!10^{-9*}$ \\
\midrule
\textbf{Mean (5 pos.)} & $\mathbf{+0.405}$ & $+0.296$ & $-0.109$ & -- \\
\textbf{Mean (all 8)}  & $+0.188$ & $+0.128$ & $-0.060$ & -- \\
\bottomrule
\end{tabular}
\end{table}

\paragraph{Failure-mode analysis of negative-selectivity cases.}
We examined all negative-gap designs under the TS-S2 seed-1024 checkpoint in a pool of 482 CaM designs from nine generator--guidance configurations. After aligning each receptor to the 3CLN holo reference, we annotated five non-exclusive categories: length below 50 residues, binder centre-of-mass more than 25\,\AA{} from the designated groove (residues 84--88 and 124--125), fewer than five binder residues within 10\,\AA{} of the receptor, an inter-chain C$_\alpha$ distance below 2.5\,\AA{}, and apo score above 0.3. Ten designs ($2.1\%$), all from PXDesign with Langevin refinement, had negative gaps (Table~\ref{tab:failure_modes_ts1024}). All ten met the length criterion, four had misplaced interfaces, four had clashes, and two had insufficient interfaces. The length criterion applies to this generated-backbone analysis. Scorer errors and experimental state preference remain unresolved by these structural annotations.
\begin{table}[htbp]
\centering
\caption{Failure-mode analysis of the $10$ negative-selectivity designs out of $482$ total designs from $9$ guidance pipelines, scored under the TS-S2 seed-1024 $Q_\theta$
checkpoint. Categories are not mutually exclusive: a single design may fall into several modes simultaneously.}
\label{tab:failure_modes_ts1024}
\begin{tabular}{lcc}
\toprule
\textbf{Failure mode} & Count & Fraction \\
\midrule
too\_short ($<50$ residues)        & 10/10 & $\mathbf{100.0\%}$ \\
wrong\_binding\_site ($>25$\,\AA)  & 4/10  & $40.0\%$ \\
steric\_clash ($<2.5$\,\AA\ inter-chain $C_\alpha$) & 4/10 & $40.0\%$ \\
insufficient\_interface ($<5$ residues at $10$\,\AA) & 2/10 & $20.0\%$ \\
apo\_binding ($Q_\text{apo} > 0.3$) & 1/10  & $10.0\%$ \\
\bottomrule
\end{tabular}
\end{table}

\paragraph{Boltz-2 cross-check.}
We compared Boltz-2 $\Delta$ipTM with $Q_\theta$ gaps on 50 vanilla designs per target (Table~\ref{tab:boltz2_validation}). Mean $\Delta$ipTM was positive on four of eight targets. Design-level rank correlations were strongest on A$_{2A}$ ($0.500$) and CaM ($0.349$), and weak or negative elsewhere. BCL-2 had a mean $\Delta$ipTM of $-0.183$ and only $24\%$ positive differences, indicating disagreement between the two scoring methods.
\begin{table}[htbp]
\centering
\caption{Boltz-2 $\Delta$ipTM as a $Q_\theta$-independent structural cross-check across 8 OOD targets ($n{=}50$ vanilla designs per target). $Q_\theta$ scores use the 3 seeds. Per-target Spearman $\rho$ is computed between $Q_\theta$ selectivity $S$ and Boltz-2 $\Delta\mathrm{ipTM} = \mathrm{ipTM}_\mathrm{holo} - \mathrm{ipTM}_\mathrm{apo}$.}
\label{tab:boltz2_validation}
\begin{tabular}{lccc}
\toprule
\textbf{Target} & Mean $\Delta$ipTM & Frac $>0$ & Spearman $\rho$ \\
\midrule
A$_{2A}$    & $+0.055$ & $50\%$ & $\mathbf{+0.500}$ \\
CaM         & $+0.032$ & $56\%$ & $\mathbf{+0.349}$ \\
MDM2        & $+0.057$ & $76\%$ & $-0.047$ \\
Ran         & $+0.045$ & $58\%$ & $+0.081$ \\
PAI-1       & $-0.022$ & $50\%$ & $+0.076$ \\
Integrin    & $-0.003$ & $50\%$ & $+0.124$ \\
BCL-2       & $-0.183$ & $24\%$ & $+0.087$ \\
ER$\alpha$  & $-0.003$ & $44\%$ & $+0.050$ \\
\bottomrule
\end{tabular}
\end{table}

\paragraph{AlphaFold-3 cross-check.}
We refolded a subset of designs with AlphaFold~3 in single-sequence mode (Table~\ref{tab:af3_iptm}). All designs for the training target ALK and held-out ER$\alpha$ had positive $\Delta$ipTM in both vanilla and refined panels. Only $5\%$ of the 19 BCL-2 designs had positive differences. The cause of this disagreement was not resolved by these measurements.
\begin{table}[htbp]
\centering
\caption{AlphaFold~3 $\Delta$ipTM in single-sequence mode. Positive values indicate holo preference. Rec.\ $\Delta$len denotes the relative length difference between binder and receptor.}
\label{tab:af3_iptm}
\begin{tabular}{lccccc}
\toprule
\textbf{Target} & \textbf{Guidance} & Mean $\Delta$ipTM & Frac.\ ${>}0$ & $n$ & Rec.\ $\Delta$len \\
\midrule
\multirow{2}{*}{ALK}
  & Vanilla  & $+0.058$ & $100\%$ & $50$ & \multirow{2}{*}{$5\%$} \\
  & Langevin & $+0.057$ & $100\%$ & $50$ & \\
\midrule
\multirow{2}{*}{ER$\alpha$}
  & Vanilla  & $+0.034$ & $100\%$ & $50$ & \multirow{2}{*}{$4\%$} \\
  & Langevin & $+0.034$ & $100\%$ & $50$ & \\
\midrule
BCL-2 & Langevin & $-0.103$ & $5\%$ & $19$ & $20\%$ \\
\bottomrule
\end{tabular}
\end{table}

\FloatBarrier
\subsection{Rosetta Interface Analysis}
\label{app:rosetta}
We computed Rosetta InterfaceAnalyzer metrics with repacked sidechains on 400 RFdiffusion designs, comprising 50 designs for each of eight targets (Table~\ref{tab:rosetta_ood}). This panel contains seven targets from the OOD test split plus the training target ALK, and excludes CaM. Mean interface energies were most favorable for BCL-2 ($-9.1$ REU) and least favorable for Ran ($25.4$ REU) and Integrin ($39.1$ REU). These energies characterize individual interfaces; a holo--apo energy comparison requires calculations in both states.

\begin{table}[htbp]
\centering
\caption{Rosetta InterfaceAnalyzer metrics for 400 vanilla RFdiffusion designs (50 per target). The panel includes seven OOD targets and training target ALK; CaM is absent.}
\label{tab:rosetta_ood}
\begin{tabular}{lrrrr}
\toprule
Target & $\Delta G$ (REU) & $\Delta$SASA (\AA$^2$) & Contacts & H-bonds \\
\midrule
BCL-2      & $-9.1$  & 968 & 40.5 & 2.3 \\
ALK        & $+0.3$  & 927 & 33.8 & 4.0 \\
MDM2       & $-1.0$  & 606 & 25.1 & 2.2 \\
A$_{2A}$   & $-0.9$  & 512 & 19.1 & 1.4 \\
ER$\alpha$ & $+3.8$  & 564 & 18.2 & 1.3 \\
PAI-1      & $+1.3$  & 464 & 16.7 & 1.4 \\
Ran        & $+25.4$ & 292 & 13.0 & 0.7 \\
Integrin   & $+39.1$ & 755 & 37.8 & 8.6 \\
\bottomrule
\end{tabular}
\end{table}

\FloatBarrier
\subsection{Statistical Analysis}
\label{app:statistical_tests}

Scoring results report means and standard deviations across three training seeds. Best-of-$K$ analysis used 10{,}000 bootstrap trials. The $p$-values in Table~\ref{tab:phase3_mpnn} are reported for paired Wilcoxon signed-rank comparisons of vanilla and refined designs. These paired tests quantify changes between methods. BLI measurements were performed in duplicate and are reported descriptively, without significance tests or confidence intervals for $K_D$.
% Author check: the original prose and table disagree on Wilcoxon sidedness and
% the target-level p-values. Preserve reported table values pending raw-data verification.

\FloatBarrier
\subsection{Additional Held-Out Target: Adenylate Kinase}
\label{app:adk}
To examine transfer beyond the benchmark, we evaluated adenylate kinase (AdK), which is outside the 65-target benchmark listed in Table~\ref{tab:split_v5}. All reported combinations had positive mean $Q_\theta$ gaps (Table~\ref{tab:design_adk}). TDS produced the largest mean gap for RFdiffusion ($0.543$) and Proteina-ComplexA ($0.904$), while SMC was highest for PXDesign ($0.311$). Boltz-2 holo--apo ipTM differences were mixed, so independent support for the predicted state preference remained inconclusive. Candidate counts vary by method. Structural and sequence metrics are listed in Table~\ref{tab:adk_structure}.

\begin{table}[htbp]
\centering
\caption{Additional AdK design results, scored by the three-seed ensemble. RF, PX, and Pro denote RFdiffusion, PXDesign, and Proteina-ComplexA. Des and SR denote computational designability and joint success rates.}
\label{tab:design_adk}
\begin{tabular}{lrrrrrr}
\toprule
Method & $N$ & $\bar S$ & $\sigma_S$ & $S>0$ & Des & SR \\
\midrule
RF/TDS & 50 & $\mathbf{0.543}$ & $\mathbf{0.108}$ & $100\%$ & $\mathbf{84.0\%}$ & $\mathbf{84.0\%}$ \\
RF/SMC & 18 & $0.522$ & $0.160$ & $100\%$ & $83.3\%$ & $83.3\%$ \\
RF/Langevin & 50 & $0.372$ & $0.262$ & $100\%$ & $82.0\%$ & $82.0\%$ \\
RF/Classifier & 50 & $0.298$ & $0.248$ & $100\%$ & $82.0\%$ & $82.0\%$ \\
RF/Vanilla & 50 & $0.260$ & $0.253$ & $100\%$ & $80.0\%$ & $80.0\%$ \\
PX/SMC & 32 & $\mathbf{0.311}$ & $0.321$ & $87.5\%$ & $43.8\%$ & $40.6\%$ \\
PX/TDS & 50 & $0.305$ & $\mathbf{0.302}$ & $\mathbf{92.0\%}$ & $56.0\%$ & $\mathbf{50.0\%}$ \\
PX/Langevin & 50 & $0.245$ & $0.332$ & $76.0\%$ & $\mathbf{60.0\%}$ & $46.0\%$ \\
PX/Vanilla & 50 & $0.223$ & $0.335$ & $76.0\%$ & $58.0\%$ & $44.0\%$ \\
PX/Classifier & 50 & $0.124$ & $0.341$ & $74.0\%$ & $54.0\%$ & $42.0\%$ \\
Pro/TDS & 50 & $\mathbf{0.904}$ & $\mathbf{0.130}$ & $\mathbf{100\%}$ & $60.0\%$ & $60.0\%$ \\
Pro/SMC & 50 & $0.878$ & $0.143$ & $\mathbf{100\%}$ & $\mathbf{66.0\%}$ & $\mathbf{66.0\%}$ \\
Pro/Langevin & 50 & $0.481$ & $0.463$ & $82.0\%$ & $34.0\%$ & $26.0\%$ \\
Pro/Classifier & 50 & $0.420$ & $0.444$ & $80.0\%$ & $28.0\%$ & $20.0\%$ \\
Pro/Vanilla & 50 & $0.272$ & $0.415$ & $72.0\%$ & $34.0\%$ & $20.0\%$ \\
\bottomrule
\end{tabular}
\end{table}
\begin{table}[htbp]
\centering
\caption{AdK structural and sequence metrics corresponding to Table~\ref{tab:design_adk}. Both holo and apo ipTM are reported to preserve disagreements with the learned selectivity scores.}
\label{tab:adk_structure}
\begin{tabular}{lrrrr}
\toprule
Method & ipTM holo & ipTM apo & scTM & NLL holo \\
\midrule
RF/TDS & $0.394$ & $0.380$ & $0.649$ & $1.073$ \\
RF/SMC & $0.368$ & $0.405$ & $0.660$ & $\mathbf{1.036}$ \\
RF/Langevin & $0.395$ & $\mathbf{0.433}$ & $0.643$ & $1.049$ \\
RF/Classifier & $0.410$ & $0.415$ & $\mathbf{0.676}$ & $1.044$ \\
RF/Vanilla & $\mathbf{0.414}$ & $0.420$ & $0.661$ & $1.042$ \\
PX/SMC & $0.298$ & $0.314$ & $0.454$ & $1.262$ \\
PX/TDS & $\mathbf{0.352}$ & $\mathbf{0.361}$ & $0.510$ & $1.288$ \\
PX/Langevin & $0.309$ & $0.336$ & $0.493$ & $1.307$ \\
PX/Vanilla & $0.326$ & $0.313$ & $\mathbf{0.521}$ & $1.318$ \\
PX/Classifier & $0.348$ & $0.328$ & $0.477$ & $\mathbf{1.255}$ \\
Pro/TDS & $0.419$ & $\mathbf{0.463}$ & $0.545$ & $1.130$ \\
Pro/SMC & $\mathbf{0.421}$ & $0.423$ & $\mathbf{0.550}$ & $1.126$ \\
Pro/Langevin & $0.416$ & $0.387$ & $0.396$ & $\mathbf{1.093}$ \\
Pro/Classifier & $0.377$ & $0.388$ & $0.400$ & $1.101$ \\
Pro/Vanilla & $0.410$ & $0.395$ & $0.400$ & $1.096$ \\
\bottomrule
\end{tabular}
\end{table}

\FloatBarrier
\section{Efficiency Analysis}
\label{app:efficiency}
\FloatBarrier
\subsection{Compute Efficiency}
\label{app:computational_cost}
\begin{table}[htbp]
\centering
\caption{Computational cost on one A100 80GB GPU. RF, PX, and Pro denote RFdiffusion, PXDesign, and Proteina-ComplexA. GPU-hours correspond to the listed candidate counts.}
\label{tab:comp_cost}
\begin{tabular}{lccccc}
\toprule
\textbf{Method} & $N$ & min/des. & GPU-hr & $\bar{S}$ & $\bar{S}$/hr \\
\midrule
RF/Vanilla       & 50 & 2.0  & 1.7  & $0.249$ & $0.146$ \\
RF/Classifier    & 50 & 2.5  & 2.1  & $0.247$ & $0.118$ \\
RF/Langevin      & 50 & 2.5  & 2.1  & $0.314$ & $0.150$ \\
RF/SMC           & 64 & 2.0  & 2.1  & $0.488$ & $0.232$ \\
RF/TDS           & 40 & 2.5  & 1.7  & $0.477$ & $0.281$ \\
\midrule
PX/Vanilla     & 50 & 3.0  & 2.5  & $0.166$ & $0.066$ \\
PX/Classifier  & 50 & 4.0  & 3.3  & $0.175$ & $0.053$ \\
PX/Langevin    & 50 & 3.5  & 2.9  & $0.124$ & $0.043$ \\
PX/SMC         & 64 & 3.0  & 3.2  & $0.371$ & $0.116$ \\
PX/TDS         & 64 & 4.0  & 4.3  & $0.419$ & $0.097$ \\
\midrule
Pro/Vanilla  & 64 & 0.05 & 0.05 & $0.213$ & $\mathbf{4.26}$ \\
Pro/Classifier & 64 & 0.08 & 0.09 & $0.249$ & $2.77$ \\
Pro/Langevin   & 64 & 0.13 & 0.14 & $0.267$ & $1.91$ \\
Pro/SMC        & 64 & 0.05 & 0.05 & $0.350$ & $\mathbf{7.00}$ \\
Pro/TDS        & 64 & 0.10 & 0.11 & $0.470$ & $\mathbf{4.27}$ \\
\bottomrule
\end{tabular}
\end{table}

\FloatBarrier
\subsection{Inference Speed}
\label{app:inference}

$Q_\theta$ achieved $2.98 \pm 0.09$\,ms per-complex latency in single-sample mode on an A100 GPU. At batch size 16, amortized throughput reached 4{,}531 complexes per second, enabling $10^5$-candidate selectivity screening with two forward passes per candidate in under 45~seconds. RFdiffusion generates one CaM binder in approximately 2\,min on an A100; PXDesign requires approximately 3\,min. The per-method timings and their different candidate counts are listed in Table~\ref{tab:comp_cost}.

In the reported cost comparison, RFdiffusion with TDS had the highest $\bar{S}$/GPU-hour within RFdiffusion ($0.281$), and PXDesign with SMC had the highest ratio within PXDesign ($0.116$). RFdiffusion refinement increased time per design from 2.0 to 2.5 minutes. The ratios reflect the listed workloads, whose candidate counts differ across methods.

\FloatBarrier
\section{Experimental Validation Details}
\label{app:wetlab}
\FloatBarrier
\subsection{Candidate Selection and Assay Protocol}
\label{sec:method_wetlab}
We selected peptides from RFdiffusion and Proteina-ComplexA generator--guidance combinations using the predicted bounded gap $\Delta_q=Q_\theta(X^1,Y)-Q_\theta(X^0,Y)$. The measured panel in Table~\ref{tab:experimental_validation} contains nine candidates, one designated lower-scoring negative control, and an additional M13 positive control. Because we selected candidates by predicted preference, quantitative calibration requires a separately sampled set. The skeletal myosin light-chain kinase M13 peptide (CPC Scientific, SIGN-010) served as a Ca$^{2+}$-dependent binding control.

\paragraph{Calmodulin preparation.}
We used recombinant human calmodulin (MilliporeSigma, catalogue 208670). Holo and apo conditions were prepared by supplementation with CaCl$_2$ and EGTA, respectively, with the other assay conditions maintained identically.

\paragraph{Bio-layer interferometry.}
Adaptyv Bio measured binding on a Gator Bio Pro instrument. Designed peptides were immobilized through a Twin-Strep tag on compatible biosensors. After loading and baseline equilibration, sensors were exposed to CaM at 0, 100, and 1000\,nM for association measurements, followed by transfer into assay buffer for dissociation measurements. Measurements were performed in duplicate (Figure~\ref{fig:cam_bli}).

\paragraph{Analysis and interpretation.}
Sensorgrams were reference-subtracted using control sensors and buffer-only wells. Association and dissociation traces were fitted to a 1:1 binding model to obtain rate constants and $K_D$. Concentration-dependent responses were classified as binding; responses below detection were recorded as NB. The reported $>1.00\,\mu$M value is a bound, not a point estimate. The $1.06\,\mu$M fit is close to the highest tested concentration and should be interpreted accordingly. We report duplicate measurements without replicate-level uncertainty estimates, an independent preparation count, or a quantified detection limit. NB therefore does not establish zero affinity or a quantitative lower bound on the holo/apo selectivity ratio.
% Author check: add full buffer composition, CaCl2/EGTA concentrations,
% temperature, loading/association/dissociation times, sensorgram files, fit QC,
% replicate structure, and the assay detection limit before final submission.

\FloatBarrier
\subsection{Complete Experimental Panel}
Five designed peptides bound holo CaM, and none showed detectable apo binding under the tested conditions (Table~\ref{tab:experimental_validation}). Four candidates with predicted gaps of $0.688$--$0.993$ had no detectable binding to either state. This combination of binding and nonbinding candidates separates computational prioritization from experimental confirmation.

\begin{table}[htbp]
\centering
\caption{Peptide sequences corresponding to Table~\ref{tab:experimental_validation}. The sequences exclude assay tags.}
\label{tab:wetlab_sequences}
\begin{tabular}{ll}
\toprule
Design & Sequence \\
\midrule
RF/V 944 & ATAAMIKTFQDVVVAAVREAREK \\
RF/V 13 & SEAFARAAAVLAKARAAK \\
Pro/SM 0273 & EGFKKLLKEALEIAK \\
RF/V 657 & KVAEQAKQWILEMLAK \\
RF/Lg 25 & EKLEALLREAGAARRAAKKAAEAA \\
Pro/V 0145 & VDEDGDGKIDLPELSALLREKIK \\
RF/Lg 330 & SELTKEILKKAMEMT \\
Pro/V 0376 & AFGAEVKTPRTRFLDVLR \\
Pro/SM 0954 & EEAARAAGLARLPRPLLLLQAL \\
Pro/V 0105 & SEIAELLRRNPEGDPETLREALAA \\
M13 & KRRWKKNFIAVSAANRFKKISSSGAL \\
\bottomrule
\end{tabular}
\end{table}
The original identifiers combine the generator and scheme prefixes with \texttt{design\_} or, for Pro/SM, \texttt{smc\_particle\_}, followed by the displayed identifier. The listed sequences are given in Table~\ref{tab:wetlab_sequences}.

\clearpage
\section{Limitations and Discussion}
\label{sec:discussion}
AlloGen connects paired conformational scoring with binder generation, with computational transfer evaluated on a target-level split and state-dependent binding measured in selected CaM designs. Its use requires paired receptor structures, and transfer across sequence-clustered families remains to be tested. Backbone-only inference also differs from training on complete complexes, and limited apo-decoy coverage may constrain performance on subtle conformational changes. Richer apo conformations and broader target panels would help assess these dependencies.

Absolute $Q_\theta$ outputs are uncalibrated scores, and reranking and guidance are evaluated partly with the scorer used for optimization. Agreement across independent computational metrics was mixed, reinforcing the role of direct binding measurements. The CaM BLI panel established binding in selected designs, but duplicate measurements without a quantified detection limit or replicate-level uncertainty cannot quantify a holo/apo selectivity ratio. Additional targets, orthogonal assays, and calibration data are needed to establish quantitative selectivity and broader experimental transfer. With this validation, we envision using AlloGen to design and prioritize binders for defined receptor conformations across protein families.
\paragraph{Broader impact.} State-selective binder design may support molecular probes and future therapeutic research. Therapeutic applications require further testing of cellular efficacy, delivery, and safety. Protein-design methods can also be misused to alter biological interactions, so downstream applications require context-specific assessment and experimental validation.

\clearpage
\section*{NeurIPS Paper Checklist}

\begin{enumerate}

\item {\bf Claims}
    \item[] Question: Do the main claims made in the abstract and introduction accurately reflect the paper's contributions and scope?
    \item[] Answer: \answerYes{} % Replace by \answerYes{}, \answerNo{}, or \answerNA{}.
    \item[] Justification: The abstract and Introduction distinguish predicted selectivity from BLI measurements on CaM. The held-out scoring benchmark, generator comparisons, and complete experimental panel support the stated scope.

\item {\bf Limitations}
    \item[] Question: Does the paper discuss the limitations of the work performed by the authors?
    \item[] Answer: \answerYes{} % Replace by \answerYes{}, \answerNo{}, or \answerNA{}.
    \item[] Justification: Appendix~\ref{sec:discussion} discusses paired-structure requirements, target-level splitting, uncalibrated scores, optimization with the evaluation scorer, mixed independent metrics, and the limited CaM assay panel.

\item {\bf Theory assumptions and proofs}
    \item[] Question: For each theoretical result, does the paper provide the full set of assumptions and a complete (and correct) proof?
    \item[] Answer: \answerNA{} % Replace by \answerYes{}, \answerNo{}, or \answerNA{}.
    \item[] Justification: The paper presents an empirical method and no formal theoretical results. The rigid-motion invariance of the geometric features is explained in Section~\ref{sec:prelim_backbone}.

    \item {\bf Experimental result reproducibility}
    \item[] Question: Does the paper fully disclose all the information needed to reproduce the main experimental results of the paper to the extent that it affects the main claims and/or conclusions of the paper (regardless of whether the code and data are provided or not)?
    \item[] Answer: \answerNo{} % Replace by \answerYes{}, \answerNo{}, or \answerNA{}.
    \item[] Justification: The architecture, objectives, and computational settings are described in Sections~\ref{sec:method} and \ref{app:implementation}. The available assay description does not specify complete buffer and timing parameters, replicate-level data, or detection limits.

\item {\bf Open access to data and code}
    \item[] Question: Does the paper provide open access to the data and code, with sufficient instructions to faithfully reproduce the main experimental results, as described in supplemental material?
    \item[] Answer: \answerNo{} % Replace by \answerYes{}, \answerNo{}, or \answerNA{}.
    \item[] Justification: The public project at \url{https://huggingface.co/ChatterjeeLab/AlloGen} includes code, model checkpoints, inference instructions, and a sample dataset. Availability of the complete benchmark and raw experimental records sufficient to reproduce every reported result has not been established.

\item {\bf Experimental setting/details}
    \item[] Question: Does the paper specify all the training and test details (e.g., data splits, hyperparameters, how they were chosen, type of optimizer) necessary to understand the results?
    \item[] Answer: \answerNo{} % Replace by \answerYes{}, \answerNo{}, or \answerNA{}.
    \item[] Justification: The paper specifies the architecture, target split, learning rates, batch sizes, dropout, and guidance settings. The description does not fully specify optimizer selection, all model-selection decisions, or every assay setting.

\item {\bf Experiment statistical significance}
    \item[] Question: Does the paper report error bars suitably and correctly defined or other appropriate information about the statistical significance of the experiments?
    \item[] Answer: \answerNo{} % Replace by \answerYes{}, \answerNo{}, or \answerNA{}.
    \item[] Justification: Main scoring results include three-seed means and standard deviations, and reranking uses bootstrap trials. Design and BLI results do not uniformly include uncertainty estimates, and the paired-test description does not fully specify sidedness or multiple-comparison correction.

\item {\bf Experiments compute resources}
    \item[] Question: For each experiment, does the paper provide sufficient information on the computer resources (type of compute workers, memory, time of execution) needed to reproduce the experiments?
    \item[] Answer: \answerNo{} % Replace by \answerYes{}, \answerNo{}, or \answerNA{}.
    \item[] Justification: Appendix~\ref{app:efficiency} reports A100 inference latency and generation costs by method. The total compute used for training, preliminary runs, and the full research project is not documented.
    
\item {\bf Code of ethics}
    \item[] Question: Does the research conducted in the paper conform, in every respect, with the NeurIPS Code of Ethics \url{https://neurips.cc/public/EthicsGuidelines}?
    \item[] Answer: \answerYes{} % Replace by \answerYes{}, \answerNo{}, or \answerNA{}.
    \item[] Justification: The work uses public protein structures and in vitro binding assays, without human participants or sensitive personal data. External methods are cited, and funding and competing interests are disclosed.

\item {\bf Broader impacts}
    \item[] Question: Does the paper discuss both potential positive societal impacts and negative societal impacts of the work performed?
    \item[] Answer: \answerYes{} % Replace by \answerYes{}, \answerNo{}, or \answerNA{}.
    \item[] Justification: Appendix~\ref{sec:discussion} discusses potential applications and the need for downstream biological validation, along with possible misuse of protein-design methods.
    
\item {\bf Safeguards}
    \item[] Question: Does the paper describe safeguards that have been put in place for responsible release of data or models that have a high risk for misuse (e.g., pre-trained language models, image generators, or scraped datasets)?
    \item[] Answer: \answerNA{} % Replace by \answerYes{}, \answerNo{}, or \answerNA{}.
    \item[] Justification: The released artifact is a specialized interface scorer and associated code, not a general-purpose generative foundation model. This work does not release a high-risk human dataset.

\item {\bf Licenses for existing assets}
    \item[] Question: Are the creators or original owners of assets (e.g., code, data, models), used in the paper, properly credited and are the license and terms of use explicitly mentioned and properly respected?
    \item[] Answer: \answerNo{} % Replace by \answerYes{}, \answerNo{}, or \answerNA{}.
    \item[] Justification: Existing methods and structural data sources are cited. The project repository lists an Apache-2.0 license, but the manuscript does not enumerate the license and usage terms of every external asset.

\item {\bf New assets}
    \item[] Question: Are new assets introduced in the paper well documented and is the documentation provided alongside the assets?
    \item[] Answer: \answerYes{} % Replace by \answerYes{}, \answerNo{}, or \answerNA{}.
    \item[] Justification: The model, data-construction procedure, and guidance methods are described in the paper. The public project documents released code, checkpoints, and sample data, without establishing availability of the full experimental dataset.

\item {\bf Crowdsourcing and research with human subjects}
    \item[] Question: For crowdsourcing experiments and research with human subjects, does the paper include the full text of instructions given to participants and screenshots, if applicable, as well as details about compensation (if any)? 
    \item[] Answer: \answerNA{} % Replace by \answerYes{}, \answerNo{}, or \answerNA{}.
    \item[] Justification: The work does not involve crowdsourcing or human participants. Experimental validation used recombinant calmodulin and designed peptides.

\item {\bf Institutional review board (IRB) approvals or equivalent for research with human subjects}
    \item[] Question: Does the paper describe potential risks incurred by study participants, whether such risks were disclosed to the subjects, and whether Institutional Review Board (IRB) approvals (or an equivalent approval/review based on the requirements of your country or institution) were obtained?
    \item[] Answer: \answerNA{} % Replace by \answerYes{}, \answerNo{}, or \answerNA{}.
    \item[] Justification: The study does not involve research with human subjects.

\item {\bf Declaration of LLM usage}
    \item[] Question: Does the paper describe the usage of LLMs if it is an important, original, or non-standard component of the core methods in this research? Note that if the LLM is used only for writing, editing, or formatting purposes and does \emph{not} impact the core methodology, scientific rigor, or originality of the research, declaration is not required.
    %this research? 
    \item[] Answer: \answerNA{} % Replace by \answerYes{}, \answerNo{}, or \answerNA{}.
    \item[] Justification: LLMs are not an original or non-standard component of the core method. ESM-2 is used as an optional pretrained protein-sequence feature encoder and is described and cited in the Method section. Writing and formatting assistance does not alter the methodology.

\end{enumerate}

\end{document}